%% file: main_new.tex
\documentclass[journal]{IEEEtran}
\usepackage{amsmath,amsfonts}
\usepackage{algorithm}            % 算法包
\usepackage{algpseudocode}        % 使用 algorithm 和 algpseudocode 组合
\usepackage{array}
\usepackage[caption=false,font=normalsize,labelfont=sf,textfont=sf]{subfig}
\usepackage{textcomp}
\usepackage{stfloats}
\usepackage{url}
\usepackage{verbatim}
\usepackage{graphicx}
\def\BibTeX{{\rm B\kern-.05em{\sc i\kern-.025em b}\kern-.08em
    T\kern-.1667em\lower.7ex\hbox{E}\kern-.125emX}}
\usepackage{balance}
\usepackage{multirow}
\usepackage{booktabs}
\usepackage{newunicodechar}
\usepackage{amssymb}  % For empty circle
\usepackage{pifont}   % For solid and half-solid circles
\usepackage{utfsym}
\usepackage{enumerate}
\usepackage{xspace}
\usepackage{amsthm}
\usepackage{algorithm}
\usepackage{algpseudocode}

\usepackage{enumitem}
\usepackage{colortbl}
\usepackage{subcaption} % 使用 subcaption 包
\usepackage{threeparttable}
\usepackage{threeparttablex} 
\usepackage{lipsum}
\usepackage{soul}
\usepackage{url}
\usepackage{hyperref}
\usepackage{tcolorbox}

\newunicodechar{✘}{\usym{2717}}
\newunicodechar{✔}{\usym{2713}}
\newunicodechar{◐}{\usym{1F313}}
\newunicodechar{◑}{\usym{1F317}}
\newunicodechar{●}{\usym{1F311}}
\newunicodechar{◯}{\usym{1F315}}

\newcommand{\malscan}{{MalScan}\xspace}
\newcommand{\mamadroid}{{MaMaDroid}\xspace}
\newcommand{\apigraph}{{APIGraph}\xspace}
\newcommand{\hrat}{{HRAT}\xspace} % CCS21
\newcommand{\bagammo}{{BagAmmo}\xspace}
\newcommand{\tool}{\textsc{FCGHunter}\xspace}

\begin{document}
\title{\tool: Towards Evaluating Robustness of Graph-Based Android Malware Detection 
% via Dependency-Aware Mutation and Multi-Objective Optimization
}
\author{
Song Shiwen,
Xiaofei Xie, 
Ruitao Feng, 
Qi Guo, 
and Sen Chen
}

\markboth{Journal of \LaTeX\ Class Files,~Vol.~18, No.~9, December~2024}%
{}

\maketitle

\begin{abstract}
% To combat the threats posed by Android malware, various Android malware detection (AMD) approaches have emerged, with graph-based methods showing particular promise due to their deeper semantic insights from Function Call Graphs (FCGs).
Graph-based detection methods leveraging Function Call Graphs (FCGs) have shown promise for Android malware detection (AMD) due to their semantic insights. However, the deployment of malware detectors in dynamic and hostile environments raises significant concerns about their robustness. While recent approaches evaluate the robustness of FCG-based detectors using adversarial attacks, their effectiveness is constrained by the vast perturbation space, particularly across diverse models and features.
To address these challenges, we introduce \tool, a novel robustness testing framework for FCG-based AMD systems. Specifically, \tool employs innovative techniques to enhance \textit{exploration} and \textit{exploitation} within this huge search space. 
Initially, it identifies critical areas within the FCG related to malware behaviors to narrow down the perturbation space. 
We then develop a dependency-aware crossover and mutation method to enhance the \textit{validity} and \textit{diversity} of perturbations, generating diverse FCGs. 
Furthermore, \tool leverages multi-objective feedback to select perturbed FCGs, significantly improving the search process with interpretation-based feature change feedback.
Extensive evaluations across 40 scenarios demonstrate that \tool achieves an average attack success rate of 87.9\%, significantly outperforming baselines by at least 44.7\%. Notably, \tool achieves a 100\% success rate on robust models (e.g., AdaBoost with \malscan), where baselines achieve only 11\% or are inapplicable.

% We conducted extensive evaluation across 40 different scenarios, incorporating eight types of features and five model variants. The results demonstrate that \tool achieves an average attack success rate of 87.9\%, which substantially outperforms baselines by at least 44.7\%. 
% Its effectiveness is particularly pronounced in testing more robust models; for instance, \tool achieves a 100\% success rate on the robust model (i.e., AdaBoost with \malscan), where baselines are either inapplicable or achieve only an 11\% success rate. 

\end{abstract}


\input{tex/1-introduction}
\input{tex/2-preliminaries}
\input{tex/3-threat_model}
\input{tex/4-challenges}
\input{tex/5-methodology}
\input{tex/6-evaluation}
\input{tex/8-limitations_and_discussion}
\input{tex/7-related_work}

\input{tex/9-conclusion}

\balance
\bibliographystyle{IEEEtran}
\bibliography{main_new}

\begin{IEEEbiography}[{\includegraphics[width=1in,height=1.25in,clip,keepaspectratio]{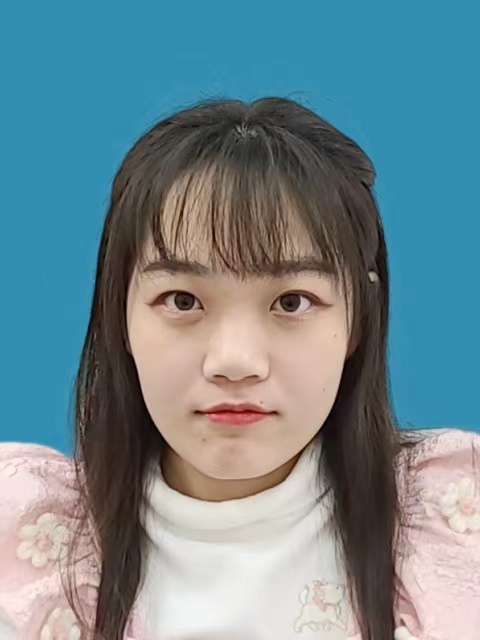}}]{Shiwen Song}
received the B.E. degree in software engineering from Henan University, China, in 2020, and the M.E. degree in software engineering from Nanjing University, China, in 2023. She is currently pursuing the Ph.D. degree with the school of computing and information systems, Singapore Management University, starting in 2024. Her research interests include malware detection, program analysis, and Android forensics.
\end{IEEEbiography}

\begin{IEEEbiography}[{\includegraphics[width=1in,height=1.25in,clip,keepaspectratio]{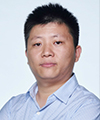}}]{Xiaofei Xie} is an Assistant Professor and Lee
Kong Chian Fellow at Singapore Management University. He obtained his Ph.D from Tianjin University
and won the CCF Outstanding Doctoral Dissertation Award (2019) in China. Previously, he was a
Wallenberg-NTU Presidential Postdoctoral Fellow at
NTU. His research mainly focuses on the quality
assurance of both traditional software and AI-enabled
software. He has published top-tier conference/journal
papers in the areas of software engineering, security
and AI, focusing on the use of AI for software testing
and the testing and security of AI systems. In particular, he has received four
ACM SIGSOFT Distinguished Paper Awards and a APSEC Best Paper Award.
\end{IEEEbiography}

\begin{IEEEbiography}[{\includegraphics[width=1in,height=1.25in,clip,keepaspectratio]{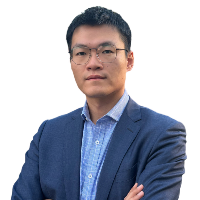}}]{Ruitao Feng}
is a Lecturer at Southern Cross University, Australia. He received the Ph.D. degree from the Nanyang Technological University. His research centers on security and quality assurance in software-enabled systems, particularly AI4Sec\&SE. This encompasses learning-based intrusion/anomaly detection, malicious behavior recognition for malware, and code vulnerability detection.
\end{IEEEbiography}

\begin{IEEEbiography}[{\includegraphics[width=1in,height=1.25in,clip,keepaspectratio]{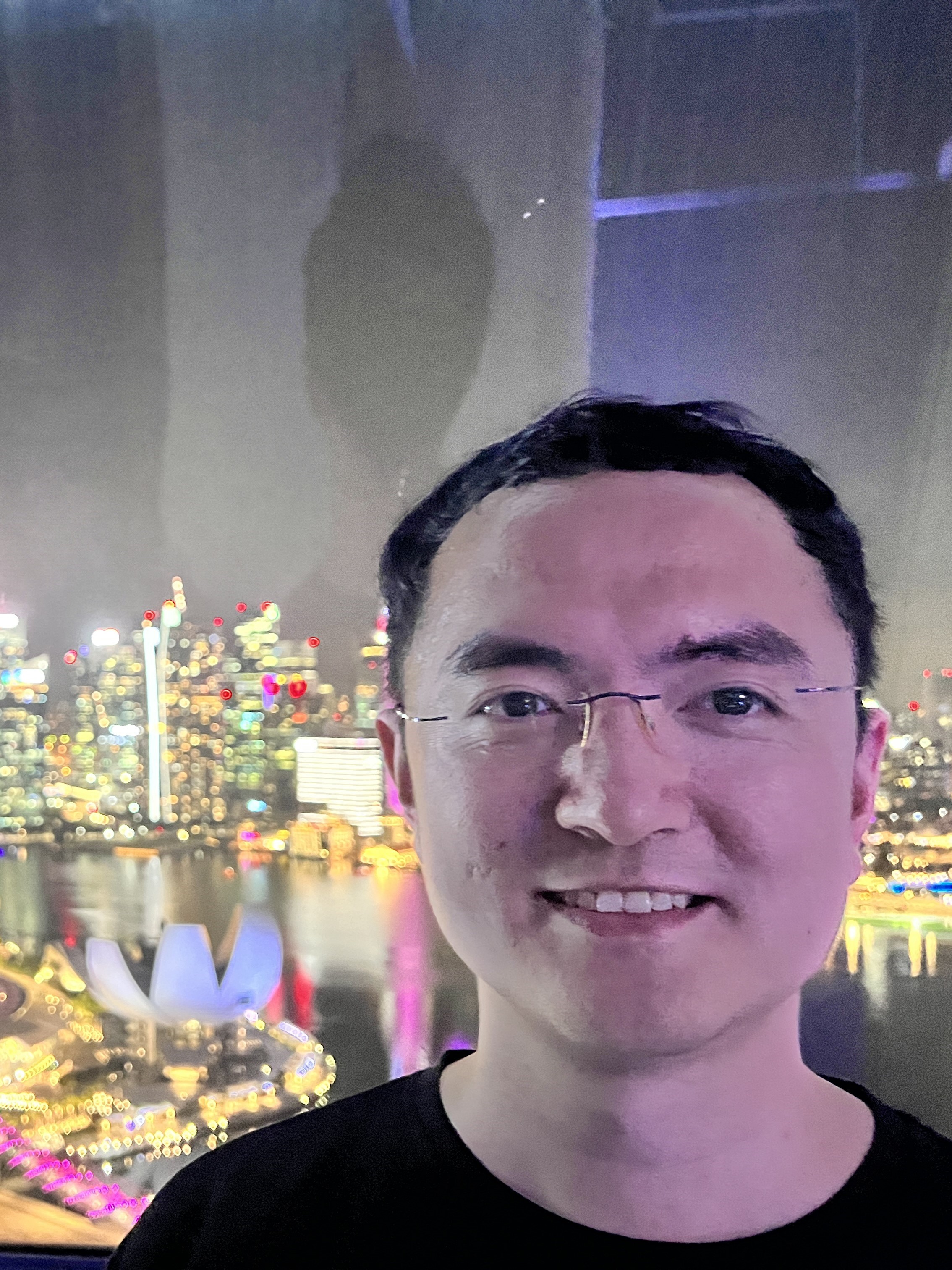}}]{Qi Guo} received the B.E. degree from Shanghai Jiao Tong University, China, in 2010, and the M.E. degree from Tianjin University, China. He is currently pursuing the Ph.D. degree with the college of intelligence and computing, Tianjin University, China, since 2019. His research interests include code retrieval and code completion. His papers have been published in top-tier software engineering conferences, such as ICSE.

\end{IEEEbiography}

\begin{IEEEbiography}[{\includegraphics[width=1in,height=1.25in,clip,keepaspectratio]{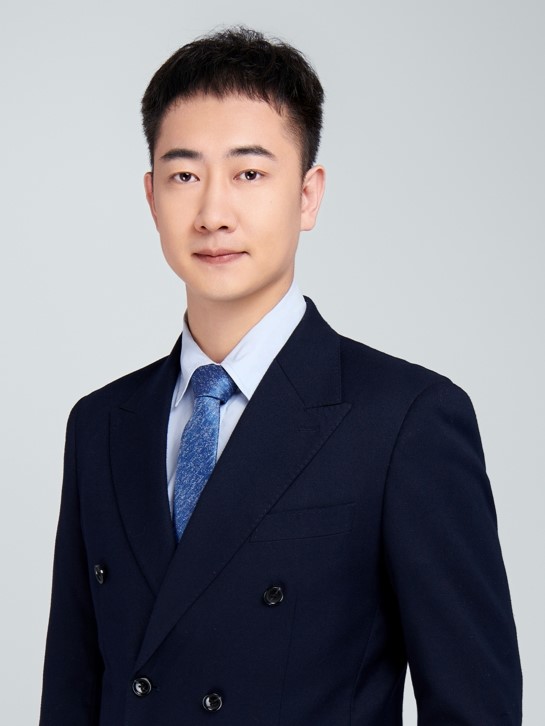}}]{Sen Chen}
 (Member, IEEE) is an Associate Professor
at the College of Intelligence and Computing, Tianjin
University, China. Before that, he was a Research
Assistant Professor at Nanyang Technological University, Singapore. His research focuses on software
and system security. He got six ACM SIGSOFT
Distinguished Paper Awards. More information is
available on \href{https://sen-chen.github.io/}.
\end{IEEEbiography}

\end{document}

%% file: tex/1-introduction.tex
% \vspace{-0.15cm}
\section{Introduction}\label{introduction}
\IEEEPARstart{A}{ndroid}
malware, such as those designed to steal users' privacy or device resources, has become a major threat to mobile security~\cite{chatterjee2018spyware, suarez2020eight, sun2021mind}. This growing threat, fueled by the popularity and openness of the Android platform, has driven the development of various detection methods. In recent years, machine learning (ML)-based approaches have been widely applied in Android malware detection (AMD), demonstrating promising results by leveraging static features of applications~\cite{arp2014drebin,garcia2018lightweight,xiao2019image,yuan2020byte,wu2019malscan,onwuzurike2019mamadroid,zhang2020enhancing}.
These methods can be mainly divided into two categories, i.e., string-based detection (e.g., Drebin~\cite{arp2014drebin}), and graph-based detection (e.g., \malscan~\cite{wu2019malscan}).
Graph-based methods have emerged as a particularly promising alternative~\cite{gao2024comprehensive}, offering superior performance compared to string-based ones~\cite{wu2019malscan,zhang2020enhancing,gao2024comprehensive}. 
Specifically, such methods use features extracted from the Function Call Graph (FCG) of an Android package kit (APK)'s smali code (i.e., the intermediate representation of an APK after compilation~\cite{hou2017hindroid}), which offers deep semantic insights into app behaviors and effectively identifies malicious patterns.

{
However, 
ML-based applications are widely recognized for their susceptibility to robustness issues
~\cite{pei2017deepxplore,xie2019deephunter,kim2019guiding,carlini2017towards,wang2021robot}, which can lead to severe consequences, particularly in safety- and security-critical contexts like autonomous driving and malware detection. 
For instance, attackers can make subtle modifications to malware, preserving its malicious intent while enabling it to evade detection. 
To address this, robust testing is essential before deploying ML models in dynamic and potentially hostile environments~\cite{barbero2022transcending}. To this end, adversarial attack methods~\cite{shahpasand2019adversarial,chen2019android, sriramanan2020guided, li2023black, sun2023mate} have been developed to rigorously evaluate model robustness. These evaluations help developers identify vulnerabilities, providing insights for improving robustness, such as retraining models with adversarial samples generated during testing~\cite{chen2020explore, mani2021defending, zhou2022adversarial}.
}

There are two main kinds of attacks in graph-based AMD: \textit{feature-level attacks} and \textit{code-level attacks}.
Feature-level attacks, which directly perturb the features of an APK (i.e., the model's input), can achieve high success rates~\cite{chen2019android, wu2019malscan, sun2023mate}. However, these perturbations often do not realistically reflect APK modifications, thereby compromising the fidelity of robustness assessments.
In contrast, code-level attacks alter the APK's smali code, indirectly changing its features used for detection. These attacks, conducted directly on the APK, are more realistic but inherently more complex due to the discontinuous nature of the perturbation space.

Recent studies have begun to explore code-level adversarial attacks~\cite{zhao2021structural,li2023black}. Essentially, these attacks involve modifying the smali code of an APK such that its FCG can be affected. \hrat~\cite{zhao2021structural}, the pioneering work, introduced a set of FCG-level perturbation operators that can be translated into semantically consistent code-level perturbations. Furthermore, a deep Q-network (DQN) is used to guide the perturbation generation. Meanwhile, 
% \bagammo~\cite{li2023black} presents a black-box attack strategy that employs the genetic algorithm to attack against a surrogate model of the targeted classifier. 
\bagammo~\cite{li2023black} employs a genetic algorithm (GA) that simulates targeted classifiers with a surrogate model and modifies the FCG by inserting non-executable code, optimizing the attack process.
Despite these advancements, their effectiveness is still limited, particularly when facing relatively robust scenarios~\cite{gao2024comprehensive} (e.g., \malscan~\cite{wu2019malscan}).
% these methods are still not very effective, particularly against robust features (e.g., Malscan~\cite{wu2019malscan}) or classifiers (e.g., Random Forest). 

The primary challenge lies in the \textit{vast perturbation space} in the APK, where potential modifications to an FCG can be infinite, complicating the search for adversarial perturbations. 
To effectively navigate this, a variety of perturbation operators is necessary for enhanced exploration, alongside precise feedback mechanisms for better exploitation.
However, current methods are often restricted to specific mutation types, such as only adding edges~\cite{li2023black}, or they perform multiple but simplistic perturbations~\cite{zhao2021structural}, limiting the generation of diverse FCGs. Concerning feedback mechanisms, existing approaches like \hrat~\cite{zhao2021structural} predominantly rely on gradient information, which is costly and unobtainable in non-differentiable ML classifiers like Random Forest or K-nearest neighbors (KNN). These challenges become more severe when AMDs employ diverse features and models. Additionally, we observe that current methods are mainly applied and evaluated only in limited scenarios (as depicted in Table~\ref{tab:baselinescope}) and fail to be effective in scenarios with more robust features~\cite{wu2019malscan, gao2024comprehensive} (e.g., \malscan) and popular models (e.g., Random Forest).

\begin{table}[!t]
\footnotesize
\centering
\caption{The Scope of Existing Adversarial Attack Methods.}
\label{tab:baselinescope}
% \vspace{-12pt} % 调整为您需要的间距
\scalebox{0.7}{
\begin{tabular}{c|c|c|c|c|c|c}
\toprule
\textbf{Tool} & \textbf{FCG-based} & \textbf{Deep Learning} & \multicolumn{2}{c|}{\textbf{Instance Algorithm}} & \multicolumn{2}{c}{\textbf{Ensemble Models}} \\
    & \textbf{AMD} & \textbf{MLP} & \textbf{KNN-1} & \textbf{KNN-3} & \textbf{Random Forest} & \textbf{AdaBoost} \\   
\midrule
        & \textbf{\malscan}           &     ◯    &    ●   &   ◯  &   ◯  & ◯ \\ 
\textbf{\hrat}    & \textbf{\mamadroid}         &     ◯    &    ●   &   ◯  &   ◯  &   ◯\\ 
        & \textbf{\apigraph}          &     ◯    &    ●  &   ◯  &   ◯  &   ◯\\ 
\midrule
        & \textbf{\malscan}           &      ◯   &    ◯   &   ◯  &   ◯  &   ◯ \\ 
\textbf{\bagammo} & \textbf{\mamadroid}         &      ●   &    ●   &   ●  &   ●   &   ●\\ 
        & \textbf{\apigraph}          &      ●   &    ●   &   ●  &   ●   &   ●\\ 
\bottomrule
\end{tabular}
}
\vskip 5pt
\textbf{Note:} (●) for full consideration and (◯) for no consideration.
%and half circles (◑) for partial consideration.

\vspace{-0.4cm}
\end{table}

Motivated by these issues, this paper introduces \tool, a testing method specifically designed to assess the robustness of FCG-based malware classifiers across various feature types and models. \tool optimizes a sequence of perturbations to the original FCG, such that the malware can bypass detection after the modifications. Specifically, \tool tackles the exploration and exploitation challenge in the vast perturbation space through several innovative strategies: 1) it narrows the search space by pinpointing critical areas of the FCG based on sensitive system APIs; 2) it incorporates diverse perturbation operators, including three novel types (e.g., \textit{Adding Long Edges}) that significantly impact FCG features for better exploration; 3) it introduces a dependency-aware mutation representation and a conflict-resolving strategy, ensuring the feasibility of the sequence of perturbations; and 4) for optimal exploitation, \tool employs a multi-objective optimization. Except for the model output feedback, a novel interpretation-based feedback, utilizing the SHAP method~\cite{SHAP2017}, is proposed to prioritize perturbations that significantly affect crucial features, thus improving the effectiveness of the whole search.

Technically, \tool is implemented within a genetic algorithm framework. 
% This allows \tool to leverage the parallel exploration capabilities of GAs to more thoroughly explore the perturbation space.
Each individual in the population is represented as a sequence of perturbations, where each gene is not just a single perturbation but a sub-sequence of dependent perturbations. These sub-sequences, containing highly interdependent perturbations, are considered together during crossover and mutation processes to ensure the validity of the generated FCG. 
%Following dependency-aware crossover and mutation, 
{Following this step,
individuals are selected based on interpretation-assisted fitness scores that evaluate the effectiveness of perturbations in evading detection. If conflicts arise, \tool resolves them by adjusting or removing the conflicting perturbations, ensuring that the best candidates are retained for further evolution.

%To avoid local optima, \tool reinitializes the population with the most successful FCG from the previous iteration, thus allowing for more effective adversarial FCG generation.

% \fei{We need results here}
To demonstrate the effectiveness of \tool, we conducted comprehensive experiments on 40 distinct target models, incorporating eight types of graph embeddings and five different ML classifiers. To the best of our knowledge, we are the first to evaluate AMD systems across such a broad and diverse range, covering all the scenarios outlined in  Table~\ref{tab:baselinescope}.
\tool achieves an average attack success rate of 87.9\% across these detection models, significantly outperforming state-of-the-art methods (i.e., \hrat and \bagammo) by at least 44.7\%.
Our experiments also confirm the usefulness of the key components in \tool. Based on the transferability of different models, we also applied \tool to evaluate the robustness of black-box models (i.e., VirusTotal) in the real world, revealing the robustness issues of such models.

% Additionally, we also apply \tool to evaluate the robustness of real-world black-box detector  and demonstrated its usefulness in practical settings.

% \feng{Didn't we remove the black-box experiments from the paper?}}

In summary, our main contributions are as follows:
\begin{itemize}[leftmargin=*]
% \vspace{-5pt}
    \item We expose the challenges presented by current approaches for attacking three widely-used ML model types: deep neural networks, k-nearest neighbors, and decision trees, each trained with distinct feature sets. Our analysis reveals that existing methods have limitations in certain scenarios, particularly regarding the models and feature types.
    % \item We design an innovative testing framework for evaluating the robustness of FCG-based AMD. The novelty of our approach lies in optimizing valid perturbation sequences with the multi-objective optimization.
    \item {We propose a novel robustness testing framework, incorporating dependency-aware mutation and multi-objective optimization, which can effectively evaluate different kinds of graph-based Android malware detectors. Our approach generates adversarial samples while preserving the malicious functionalities of the malware, leveraging diverse perturbation operators for enhanced exploration and precise feedback mechanisms for optimal exploitation.}
    % \item We found that tree-based models using \malscan are more robust than others. We specifically developed a constraint-aware testing strategy tailored for tree-based classifiers, enhancing the effectiveness.
    \item We conduct comprehensive experiments across 40 target models, spanning five distinct model and eight feature sets, which demonstrate the effectiveness of \tool. We have made our dataset and code publicly available~\cite{ourcode}.

\end{itemize}

%% file: tex/2-preliminaries.tex
% \vspace{-10pt}
\section{Graph-based Android Malware Detection}\label{preliminaries}
% \vspace{-5pt}
% \feng{Need a brief. Why should the following be selected as the targeted models? Currently necessary evaluation criteria (existing)/unsolved problems (what is new) in adv for FCG (ensure the readers' focus never gets lost).}
% In this section, we introduce essential preliminary knowledge covering graph-based feature extraction, machine learning based malware detectors, and existing FCG-based adversarial attacks.
% \vspace{-0.15cm}
% \subsection{FCG-based Android Malware Detection}
% \vspace{-0.15cm}
{
Graph-based detection leverages features extracted from the FCG of an APK's smali code (i.e., the intermediate representation of an APK after compilation~\cite{hou2017hindroid}), which captures the runtime behavior semantics of the application. The FCG is then transformed into a feature vector via graph embedding, which is subsequently used for binary classification to determine whether the application exhibits malicious behaviors. Next, we will introduce the main FCG-based methods, which include three graph embedding techniques and three widely used ML-based classifier types.
% Building on evaluations~\cite{wu2019malscan, onwuzurike2019mamadroid, gao2024comprehensive, yang2024novel} and adversarial attacks~\cite{chen2019android, zhao2021structural, li2023black, he2023efficient}, we highlight key FCG-based methods, including three graph embedding techniques and five widely used ML-based classifiers, which stand out due to their efficacy and adoption.
}
% Graph-based detection includes graph embedding and binary classification.
% In an FCG, nodes represent functions or abstract entities (e.g., packages or families), and edges depict call relationships. In AMD, the FCG is transformed into a feature vector via graph embedding and used as input for binary classification.

% Graph-based detection involves graph embedding and binary classification. First, the FCG of an APK’s smali code is created, with nodes representing functions or abstract entities (e.g., packages or families) and edges depicting call relationships. The FCG is then transformed into a feature vector via graph embedding, which is used for binary classification to determine if the app contains malicious patterns.

% }

\vspace{-10pt}
\subsection{Graph Embedding Methods}\label{preliminaries:embedding}
% \vspace{-5pt}
This step involves deriving a vector from an APK’s FCG, where nodes represent functions or abstract entities and edges depict call relationships, to capture crucial structural and behavioral patterns for classification. In the following, we will briefly introduce the three commonly used features~\cite{gao2024comprehensive}, i.e., \malscan, \mamadroid, and \apigraph.

%: \malscan, \mamadroid, and \apigraph.

\textbf{\malscan}~\cite{wu2019malscan} emphasizes the importance of 21,986 critical system API calls within function-level FCGs.
It employs four centrality metrics: \textit{Degree}, \textit{Katz}, \textit{Closeness}, and \textit{Harmonic}, each offering unique insights into a node's significance, and two combined centrality metrics: \textit{Average} and \textit{Concentrate}, to enhance the robustness of the extracted features.

% It utilizes six centrality metrics—\textit{Degree}, \textit{Katz}, \textit{Closeness}, \textit{Harmonic}, \textit{Average}, and \textit{Concentrate}—to assess and quantify the significance of these nodes, providing unique insights into each metric’s impact on feature robustness.

% utilizes function-level FCGs, emphasizing the importance of 21,986 critical API calls and employs six centrality metrics to  provide unique insights into their significance.
%analyzes FCGs extracted from APK files, emphasizing the importance of nodes that represent 21,986 critical system API calls.
%It identifies and quantifies the centrality of 21,986 sensitive nodes.

%malscan利用function-level的fcg，强调了21986个关键api call的重要性，使用6种中心度方法来衡量它们的重要程度。

% It employs four centrality metrics: \textit{Degree}, \textit{Katz}, \textit{Closeness}, and \textit{Harmonic}, each offering unique insights into a node's significance, and two combined centrality metrics: \textit{Average} and \textit{Concentrate}, to enhance the robustness of the extracted features.

%mamadroid model focus on call probabilities based on a Markov chain within abstract FCGs. Due to every function-level node can be abstracted to a family or a package, it has two mode 
%i.e., family-level FCGs and package-level FCG.
%So it defines 11 families and 446 families.

{\textbf{\mamadroid}~\cite{onwuzurike2019mamadroid} 
employs a Markov chain, represented as a transition matrix, to model call probabilities between abstract states (e.g., family names) within function-level FCGs, effectively characterizing app behavior. Consequently, it provides two graph embedding modes: family-level with 11 states and package-level with 446 states.}

% \textbf{\mamadroid}~\cite{onwuzurike2019mamadroid} employs a Markov chain to model transition probabilities between abstract states, which are derived from family or package names within the FCG, to characterize app behavior. \mamadroid provides two modes of feature extraction based on different levels of abstraction: family-level mode with 11 distinct states and package-level mode encompassing 446 states.

%apigraph利用android api document构建了一个知识图谱，然后使用一种基于语义的聚类方法对其他graph embedding的方法进行抽象升级，这可以大大减少特征的维度。

{\textbf{\apigraph}~\cite{zhang2020enhancing} utilizes a knowledge graph built upon the official Android API documentation to group APIs with similar functionalities or usage contexts through clustering. Therefore, it not only abstracts the representation of FCGs but also significantly reduces feature dimensions. For instance, it can reduce the dimensions in \mamadroid's package mode.}

% \textbf{\apigraph}~\cite{zhang2020enhancing} uses a graph embedding approach that extracts relationships between APIs based on a comprehensive knowledge graph built upon the official Android API documentation. Through clustering, APIs with similar functionalities or being used in similar contexts will be grouped in the FCG.  Therefore, it will replace individual functions with the clusters they belong to, which not only achieves a higher level of abstraction but also facilitates a significant reduction in feature dimension. 

% \ssw{More explanations are available on our website~\cite{ourwebsite}.} \fei{Why we need this setence here??}

% \vspace{-0.15cm}
\vspace{-10pt}
\subsection{ML-based Classifiers}\label{preliminaries:ml_dl_methods}
% \vspace{-3pt}
% \subsubsection{ML-based Classifiers in AMD}\label{preliminaries:ml_dl_methods}
% ML-based methods have been commonly applied in AMD. 
{After obtaining feature vectors via the graph embedding, ML-based methods are employed for the binary classification (i.e., malware or not).
There are three commonly used types of classifiers: deep learning (DL), instance-based learning, and ensemble-based learning.}
%i.e., MLP, KNNs with $k = 1$ and $k = 3$, and ensemble models like Random Forest and AdaBoost.

% \textbf{Multi-Layer Perceptron.}
{\textbf{Deep learning.}
Multi-Layer Perceptron (MLP) is a basic DL model widely used in AMD and adversarial attacks~\cite{yuan2019lightweight, li2021backdoor, he2023efficient,li2023black}.
MLP learns nonlinear relationships between input features and output class labels, and it outputs a continuous score from 0 to 1 that indicates the probability of a sample being malicious, using a sigmoid activation function.
% Therefore, it is susceptible to gradient-based attacks~\cite{ren2020adversarial}.
}

\textbf{Instance-based learning.} 
% \textbf{KNNs.}
The KNN algorithm, a typical instance-based method commonly used in AMD~\cite{kakavand2018application, chen2019android, zhao2021structural, he2023efficient,li2023black}, classifies data points by measuring distances (typically using Euclidean~\cite{danielsson1980euclidean} metrics) to the nearest training samples. 
For each query, it selects the k closest samples (e.g., $k=1$) and assigns a class based on the majority label among these neighbors.
% thus outputting discrete labels that indicate whether samples are malicious, potentially enhancing its robustness against attacks compared to MLP.}

% As a typical instance-based method, the KNN classifies data points by measuring distances (e.g., Euclidean distance~\cite{danielsson1980euclidean}) between them, which is popular in AMD~\cite{kakavand2018application, chen2019android, zhao2021structural, he2023efficient,li2023black}. 
% Specifically, it uses metrics like Euclidean distance~\cite{danielsson1980euclidean} to measure the distance between a given sample and all training samples. It then selects the k (e.g., $k=1$) nearest training samples and predicts the new sample’s class based on a majority vote from these neighbors' known categories. Thus, the KNN outputs discrete labels indicating whether samples are malicious.

% The KNN algorithm, a typical instance-based method used in AMD, classifies data points by measuring distances—typically using Euclidean metrics—to the nearest training samples. For each query, KNN selects the k closest samples (e.g., \( k=1 \)) and assigns a class based on the majority label among these neighbors, thus outputting discrete labels that indicate whether samples are malicious~\cite{kakavand2018application, chen2019android, zhao2021structural, he2023efficient, li2023black, danielsson1980euclidean}.

\textbf{Ensemble-based learning.} 
% \textbf{Ensemble models.} 
Random Forest and AdaBoost~\cite{chen2019android, onwuzurike2019mamadroid, he2023efficient, li2023black} effectively combine multiple learning algorithms to enhance both performance and robustness. 
Random Forest, an ensemble of decision trees, consolidates decisions through majority voting,
% This method ensures model stability by averaging predictions from trees trained independently, 
thus mitigating the influence of any single, potentially biased model.
AdaBoost sequentially applies a series of weak learners to progressively modified datasets,  thereby incrementally improving the performance of initially weak classifiers.

%% file: tex/3-threat_model.tex
\vspace{-10pt}
\section{Problem Formulation}
\vspace{-0.08cm}
{Given a target AMD system represented by model \( M(\cdot) \),} which classifies input APKs as either benign or malware, {we use $G$ and $E$ to represent the functions that extract FCG from the APK and calculate the embedding of the FCG, respectively. For a given malware $m$, the problem of AMD testing is to calculate the FCG perturbations $\delta \in \Delta$ such that:}
% \vspace{-5pt}
\begin{equation}
    % \vspace{-5pt}
    \footnotesize
    M(E(G(m))) \neq M(E(G(m)+\delta)) \wedge F(m) =F(m+reverse(\delta))
    % \vspace{-3pt}
\end{equation}
% \vspace{-2pt}
where $\Delta$ represents all possible perturbations on the vast space of FCG and $F$ represents the functionalities of the APK. 
% The perturbation must be reversible, allowing it to be mirrored at the code level (i.e., $m+reverse(\delta)$). 
The formulation sets forth three critical requirements for calculating the perturbation: 1) the perturbation should be reversible, allowing it to be mirrored at the {smali code level} (i.e., $m+reverse(\delta)$); 2) the perturbation does not affect the functionality; and 3) the feature should be alerted sufficiently to change the final prediction outcome. Addressing this problem necessitates an effective optimization-based method to search and apply these perturbations effectively.

%% file: tex/5-methodology.tex
\vspace{-5pt}
\section{Overview of \tool}\label{methodology:overview}

% \vspace{-4pt}
\begin{figure*}
  \centering
  \includegraphics[width=\linewidth]{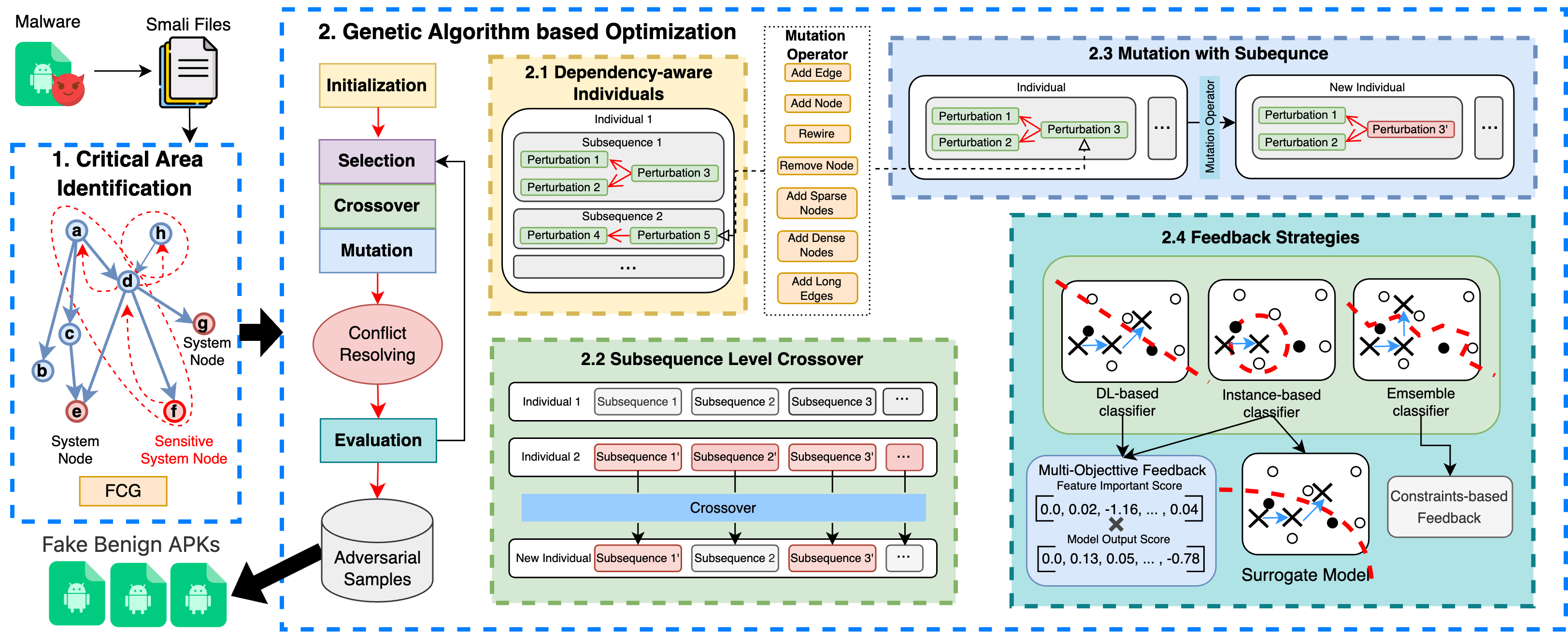}
  % \vspace{-10pt}
  \caption{Overview of \tool.}
  \label{fig:overview}
  \vspace{-10pt}
\end{figure*}

Figure~\ref{fig:overview} illustrates the main workflow of \tool, which includes identifying critical areas of the FCG and optimizing perturbations within these areas using a GA.

{\textbf{Step 1}}: Initially, \tool extracts the FCG from the malware's smali files. Given the challenge of navigating the vast perturbation space within an FCG, we first identify the critical area relevant to malware behaviors for {effectively reducing the vast perturbation space.} 

%Specifically, since malware behavior often relies on sensitive system APIs, \tool performs a reverse traversal from these nodes. This process identifies the critical area, including nodes and edges that impact these sensitive nodes, thus significantly influencing model predictions and effectively reducing the vast perturbation space.

% Recognizing the \ssw{\textit{main challenge}}, 
%of the vast potential perturbation space arbitrary modifications are deemed ineffective. 

%(\textbf{Step 1}).

% With the critical area defined, \textbf{Step 2} involves deploying the GA to optimize perturbations focused on this region. 
\textbf{Step 2}: \tool employs a GA to optimize perturbations in the identified critical area. For better exploration in the perturbation space, we incorporate seven semantics-preserving mutation operators on FCGs. Note that these mutation operators can be translated to code-level mutation that does not affect the original functionality.
% including four from previous studies \cite{zhao2021structural} and three newly developed ones.
The optimization aims to identify a \textit{sequence} of operators that orderly perturbs the FCG. Each individual in the GA population represents a perturbation sequence, enabling the mutation of diverse FCGs.
% \ssw{
% which consists of various combinations generated by the seven operators, enhancing the diversity of the new FCGs.
% The use of seven distinct mutation operators, along with their combinations in a sequence, significantly enhances the diversity of the newly generated FCGs.

%由于GA的并行探索，直接应用这些sequence在FCG上可能是不可行的。

% The direct application of crossover and mutation to this sequence may compromise the validity of perturbations. 

\begin{itemize}[leftmargin=*]
    \item {Step 2.1}: However, directly applying crossover and mutation at the level of perturbation operators to generate offspring may lead to \textit{invalid} perturbations. For example, an \textit{Add Edge} operator may become infeasible if its prerequisite node has been removed by an earlier operator within the same sequence. 
To address this issue, we perform a dependency analysis and group dependent operations into sub-sequences, ensuring the validity of the perturbation sequence.

\item {Steps 2.2 and 2.3}: Crossover and mutation processes are then performed at the level of sub-sequences to ensure the dependency. 
{Additionally, we propose a conflict-resolving mechanism to address any conflicts within a sequence after crossover and mutation.}

\item  {Step 2.4}:
% We then evaluate the new individuals and select superior offspring.
{The new individuals are evaluated to calculate their fitness, selecting the best candidates for the next iteration or stopping if optimal conditions are met in the current iteration.}
% A challenge lies in the design of the fitness function, especially considering the different model types (Challenge \textit{C3}).
%为了获取更细粒度的反馈for解决the main challenge，我们考虑了不同模型的类型，然后设计了模型特定的fitness function。
{To obtain more useful feedback, we design model-specific and explanation-based fitness functions:}
% We design model-specific fitness functions: 
a multi-objective score for MLP classifiers, a surrogate model approach for instance-based classifiers, and a constraint-based solution for decision tree classifiers. 
If a perturbation sequence successfully bypasses the target model when applied to the FCG, it is recorded as a failure test (i.e., an adversarial sample). The sequence is finally applied to alter the APK's smali code, resulting in a malware that can be misclassified as ``benign''.
\end{itemize}

%ensuring that each operation's feasibility is preserved. 

% Additionally, we propose a conflict-resolving mechanism after this process if any conflicts still within a sequence.
% After the crossover and mutation,
% Then, we evaluate the new individuals, selecting superior offspring based on their evaluation. 

% \vspace{-0.15cm}
\section{Step 1: Critical Area Identification}\label{methodology:critical_area_identification}
% \vspace{-0.15cm}
To mitigate the issue of search space explosion, we propose a specialized critical area identification method designed for FCG-based embeddings.
This method efficiently pinpoints nodes and edges sensitive to perturbations that have notable impacts on detector outcomes, {thereby reducing the search space during GA optimization. }

An FCG is obtained through static analysis of the smali code from the decompiled APK. 
% Nodes in the FCG can be categorized as system nodes and user nodes, corresponding to SDK-defined functions (i.e., APIs) and user-defined functions, respectively. 
{Nodes in the FCG are categorized as system nodes (SDK-defined functions, i.e., APIs) and user nodes (user-defined functions).}
%许多调用关系会发生在FCG中，但是用户点只能调用用户点或者系统点，而系统点只能被调用。恶意软件通常调用敏感的系统APIs来达到恶意的目的，因此FCG-based AMD通常只关注系统点。而为了保持功能不变，我们扰动的只能是用户点。因此，为了定位到受敏感系统节点影响的特征，我们的扰动方法从识别恶意软件FCG中的这些节点开始，通过匹配过程精确定位包含敏感API的节点。然后，我们从这些节点进行反向遍历，识别前面的节点和边，将这些连接区域定义为关键区域。在后续的基于GA的优化过程中，这些区域被优先处理，以确保有效的扰动。
% \ssw{Calls typically occur between user nodes or from user nodes to system nodes, while system nodes can call each other but cannot call user nodes.}
% Calls typically occur between system nodes and user nodes, with user nodes capable of calling system nodes unidirectionally.
Malware often invokes some sensitive system APIs to achieve malicious objectives (e.g., accessing user contacts). 
{In other words, malicious calls typically occur from user nodes to system nodes.}
{Therefore, FCG-based AMD typically prioritizes the user function calls that can invoke system APIs.}
% \fei{check the following sentence, not easy to understand} 
%为了定位那些可以调用关键system APIs的user function call，我们首先在图中match那些代表关键的system APIs的nodes，然后从这些nodes开始逆向的遍历整个FCG。

{To locate user function calls that can invoke critical system APIs (e.g., 21,986 sensitive APIs in \malscan and 11 family states in \mamadroid)}, we first identify the nodes representing these critical system APIs in the graph, and then perform a backward traversal from these nodes to identify preceding nodes and edges, defining these connected regions as the \textit{critical area}. Note that the perturbation can only be performed in the user functions.
% \ssw{To identify critical system nodes that can influence features ,
% we identify these nodes within the malware's FCG through a matching process, precisely pinpointing nodes that contain sensitive APIs. 
% We then perform a backward traversal from these nodes to identify preceding nodes and edges, defining these connected regions as the \textit{critical area}.
This area will be used for the subsequent GA-based optimization process.
% The area is prioritized in our subsequent GA-based optimization process.
% \vspace{-0.25cm}
\section{Step 2: Perturbation Optimization}
% \vspace{-0.15cm}
\subsection{Basic Perturbation Operators}\label{sec:perturbation}
% \vspace{-0.1cm}
Given an FCG $G=(V, E$), the nodes $V$ include both system nodes
 %(i.e., system APIs $V_s$) 
$V_s$
and user nodes 
%(i.e., user-defined functions $V_u$)
$V_u$
and an edge $(v_1, v_2)\in E$ shows the calling relationship between the two functions (with the caller $v_1$ and the callee $v_2$). In an FCG, user nodes can act as callers or callees, while system nodes can only serve as callees.
To modify the FCG, we integrate seven semantic-preserving and code-level perturbation operators. { 
The first four operators are based on prior work~\cite{zhao2021structural}, and we briefly introduce these four operators:}

% have been proven to maintain the functionality of an APK based on rules for mapping graph modifications back to smali code. More details in the paper \cite{zhao2021structural}.
% \vspace{-0.15cm}
\begin{itemize}[leftmargin=*]
    % \item \textit{Add Node:} This operator involves creating a new function $i$ and selecting a user node $a\in V_u$ to invoke $i$. Consequently, a new node $i$ and new edge $(a,i)$ are added to $G$. The new function $i$ is designed to perform non-functional operations (e.g., basic mathematical calculations) to ensure that it does not affect the overall functionality.
    \item \textit{Add Node:} This operator creates a new function $i$ and selects a user node $a\in V_u$ to invoke $i$. Consequently, a new node $i$ and new edge $(a,i)$ are added to $G$. 
    The new function $i$ is designed to perform non-functional operations (e.g., basic mathematical calculations) to ensure that it does not affect the overall functionality.
    \item \textit{Add Edge:} This operator establishes a new calling between two existing functions,  $i\in V_u$ and $f\in V$, resulting in a new edge $(i, f)$  within $G$. 
    To maintain original functionality, strategies such as using \textit{try-catch} blocks~\cite{li2023black} and unreachable conditions~\cite{zhao2021structural} ensure that the callee $f$ is never invoked, even though the calling is in the $G$.
    % \item \textit{Rewire:} This operator removes an existing edge $(a, d)$ and selects a user node $h\in V_u$  as an intermediary to reconnect  $(a, d)$. This adds two new edges: $(a, h)$ and $(h, d)$, where $(a, h)$ indicates the caller $a$'s invocation to intermediary $h$, and $(h, d)$ signifies $h$'s invocation to the original callee $d$. Special branches are added to related functions to ensure that original invocations of $d$ and $h$ remain unaffected.
    \item \textit{Rewire:} This operator removes an existing edge $(a, d)$ and selects a user node $h \in V_u$ as an intermediary, adding two new edges: $(a, h)$ and $(h, d)$, where $(a, h)$ is $a$'s invocation to $h$, and $(h, d)$ is $h$'s invocation to $d$. 
    Special branches are added to related functions to ensure the original invocations of $a$ and $d$ remain unaffected.
    % \item \textit{Remove Node:}  This targets a user node $d\in V_u$ for removal. It identifies all original callers  $\{h|(h, d)\in E\}$ and replaces the invocation statements with $d$'s function body. Subsequently, node $d$ and its connecting edges  $\{(h,d)\in E|h \in V\}$ are removed. New edges $\{(h,v)| (h,d)\in E \wedge (d,v)\in E\}$  are added to the FCG, where $h$ is the original caller of $d$ and $v$ is its callee.
    \item \textit{Remove Node:}  This operator removes a user node $d\in V_u$. 
    For maintain functionality, it identifies all original callers  $\{h|(h, d)\in E\}$ and replaces the invocation statements with $d$'s function body in the code. 
    Correspondingly, the edges  $\{(h,d)\in E|h \in V_u\}$ are removed, and new edges $\{(h,v)| (h,d)\in E \wedge (d,v)\in E\}$  are added to the $G$, where $h$ are the original callers of $d$ and $v$ are its callees.
\end{itemize}

However, these operators are very basic and insufficient for modifying features, particularly those in robust models (e.g., \malscan, which is sparser than others), often leading the GA toward local optima. Therefore, we introduce three new perturbation operators designed to substantially affect features:

% for particularly robust models,

% For instance, in an FCG that contains only one sensitive node, basic operators such as \textit{AddNode} might not adequately influence these sparse features. 
% Moreover, in such ensemble models, it's challenging to construct influential patterns from basic operators due to GA's random nature. 
% This can lead the GA to converge prematurely to local optima, ultimately failing to alter the model’s predictions. 

% \begin{figure}[!t]
%   \centering
% \includegraphics[width=0.8\linewidth]{figure/operator_new.png}
%   \caption{Perturbation Operators}
%   \label{fig:mutation_operators}
% \end{figure}

\begin{itemize}[leftmargin=*]
    \item \textit{Add Sparse Nodes:}
    {This operator adds $k$ nodes ${v_1, v_2, \ldots, v_k}$ to the $G$ \textit{at once}.
    To affect the area around an existing node $a \in V_u$, edges  
    $(a, v_1), (a, v_2), \ldots, (a, v_k)$ are added. 
    This dilutes the centrality of other nodes and redistributes the influence across the $G$.}
    \item \textit{Add Dense Nodes:}
    {This operator first performs the \textit{Add Sparse Nodes} operator, then adds new edges $\{(v_i, v_j)| i< j \wedge i,j\in[1,k]\}$ to the $G$.
    This effectively creates a dense subgraph, decreasing the relative importance of other paths in the $G$, which is particularly impactful for path-based analysis methods (e.g., Katz in \malscan).}
    \item \textit{Add Long Edges:} 
    {This operator adds $m$ long edges between two existing nodes $a \in V_u$ and $f\in V_s$. For each long edge, $k$ new nodes ${v_1, v_2, \ldots, v_k}$ are added sequentially, creating the edges  
    $(a, v_1), (v_1, v_2), \ldots, (v_{k-1}, v_k)$, and finally an edge $(v_k, f)$ to create a path between $a$ and $f$.
    This increases the number of paths leading to $f$ in $G$, significantly boosting its centrality in the network.}
    % This significantly boosts the importance of the $f$ in the $G$ by increasing the number of paths leading to it, thereby enhancing its centrality in the overall network.
\end{itemize}
These three operators are based on the combinations of four basic operators, thus still ensuring the functional integrity of the APK.
Differently, they introduce a greater magnitude of perturbation for affecting the graph's features by allowing for the adjustment of parameters (e.g., $k$), which increases the population's diversity and helps avoid the risk of GA falling into local optima (see results in \S \ref{evaluation:RQ3 ablation}).
% Therefore, the three new operators are effective in the more robust ensemble models (i.e., Random Forest and AdaBoost) (details in \S \ref{evaluation:RQ3 ablation}).}

\textbf{Translating FCG-Based Mutation to Code-Level Perturbation.} It is essential to convert FCG-based mutations into code-level modifications. These modifications should be repackaged into an APK that retains the same functionalities as the original. Specifically, the mutation in the FCG can be mapped to corresponding changes in the code as follows\footnote{More detailed illustration and code change examples can be found in~\cite{zhao2021structural}.}:

\begin{itemize}[leftmargin=*]
    \item \textit{Add Node}: We introduce a new function (i.e., node $i$) in the code that does not affect the original functionality (e.g., only printing or simple calculation like \( int j = j + 1 \), and then returns \( j \). The existing function (i.e., the user node $a$ in FCG) is modified to call this new function, but it does not process or utilize any of the returned value, thereby preserving the semantics of original function.

    \item \textit{Add Edge}: We add an invocation from a user function \(a\) to any other function \(b\). To guarantee the functionality of original function \(a\), we can prevent the actual execution of function \(b\) by introducing a \(condition\) parameter in \(b\) and insert an \(if\)-\(else\) statement in its function body. When the function \(a\) invokes \(b\), \(condition\) is set to \(true\), causing \(b\) to return a value immediately, without executing the original logic in \(b\). For invocations from \(b\)'s original callers, \(condition\) is set to \(false\), allowing \(b\) to execute its original logic, thereby preserving the original functions.

    \item \textit{Rewire}: We replace an existing call from function \(a\) to function \(c\) with an intermediary function \(b\), so that the call flow becomes \(a \rightarrow b \rightarrow c\).
    To achieve this, we replace \(a\)’s call to \(c\) with a call to \(b\) and add a call to \(c\) within \(b\). To ensure \(b\)’s original callers remain unaffected, we apply a strategy similar to \textit{Add Edge}, i.e., using a \(condition\) parameter.
    
    % \(b\) is modified to include a \(condition\) parameter and an \(if\)-\(else\) statement. If the call to \(b\) originates from \(a\), \(condition\) is set to \(true\), triggering the call to \(c\). For calls from \(b\)’s other original callers, \(condition\) is set to \(false\), allowing \(b\) to execute its original logic.

    \item \textit{Remove Node}: We delete function \(a\), which results in the removal of all calling relationships involving \(a\) in the original graph. To ensure that the program logic remains unaffected, we copy \(a\)'s function body into all its caller functions as an inline code implementation. Consequently, in the final graph, direct connections are established between \(a\)'s original callers and its callees.

    \item \textit{Add Sparse Nodes}:
    We insert \(k\) functions simultaneously, all of which are called by a single existing function \(a\). To maintain original program semantics, similar operations as in the \textit{Add Node} process are applied.

    \item \textit{Add Dense Nodes}: 
    We start by performing the same operation as in \textit{Add Sparse Nodes}. Then, for the newly added \(k\) functions, we sequentially connect them with calls. Throughout this process, we apply the same method as in the \textit{Add Edge} operation to ensure that program semantics remain unchanged.

    \item \textit{Add Long Edges}: 
    Suppose we only insert a long edge, we insert $k$ intermediate functions between an existing function \(a\) and function \(c\), creating a nested call sequence. Essentially, this establishes a chain of function calls, where \(a\) calls the first intermediate function, which in turn calls the next, and so on, until reaching \(c\). These intermediate $k$ functions are newly added and serve solely as proxies, relaying the call from \(a\) to \(c\) without affecting any other existing functions or altering the program’s original functionality.
\end{itemize}

Note that our approach mainly utilizes FCG-based mutations instead of arbitrary direct code mutations (e.g., transforming \(m=m*2\) to \(m = m << 1\)). This is because we focus on FCG-based AMDs that rely solely on the features of the FCG. Arbitrary code mutations may not always impact the FCG, and therefore, might not effectively influence robustness.

% \vspace{-0.35cm}
\subsection{Step 2.1: Individual Representation}\label{methodology:individual_representation}
% \vspace{-0.15cm}
To initialize the GA's population, a simple way is to initialize each individual as a sequence of perturbation operators ($o_1, o_2, \ldots, o_n$), where $o_i$ denotes one of the seven available perturbation operators. The sequence is then applied to the FCG $G$ to generate a new FCG $G'$.

\begin{figure}[!t]
  \centering
  % \vspace{-0.1cm}
  \includegraphics[width=0.8\linewidth]{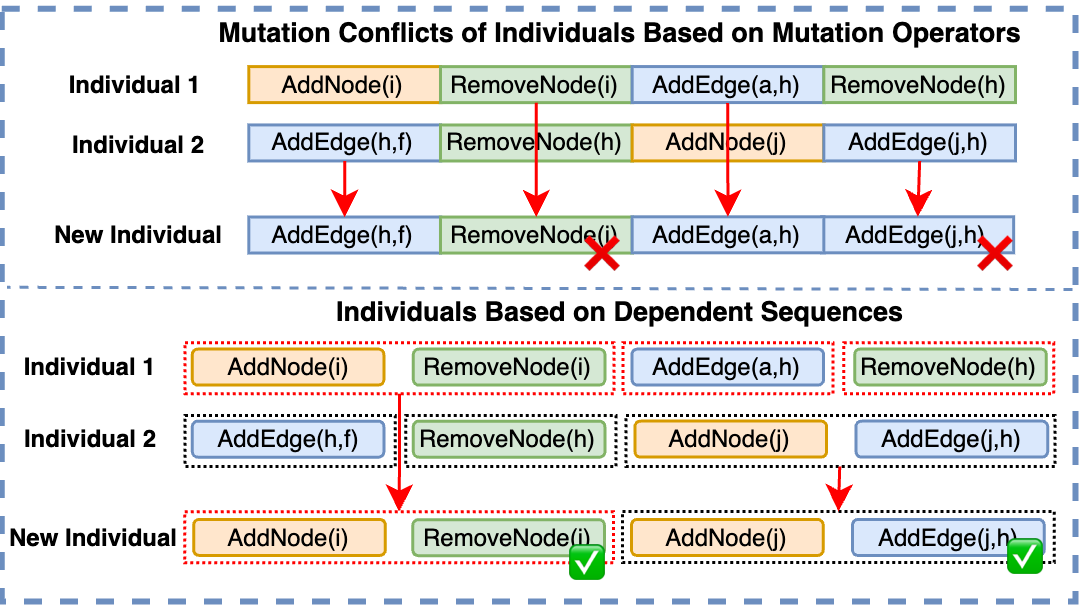}
  % \vspace{-0.2cm}
  \caption{Crossover with/without Dependency Strategy.}
  \label{fig:dependency_example}
  \vspace{-0.4cm}
\end{figure}

However, dependencies among operations often result in the generation of invalid perturbation sequences during crossover and mutation processes. {For example, as shown in the top of Figure~\ref{fig:dependency_example}, two sequences of individuals undergo crossover, creating an infeasible sequence in the new individual. Without the $AddNode(i)$ operator, the $RemoveNode(i)$ operator becomes infeasible as the node $i$ does not exist in the $G$. This issue can also arise during mutation, which greatly affects the testing efficiency.

{To address this issue, we propose a dependency-aware representation to avoid operator conflicts during crossover and mutation processes.
The approach involves a preliminary dependency analysis, grouping dependent operations into sub-sequences. Then crossover and mutation are performed at the level of sub-sequence.}

% (see the bottom part in Figure~\ref{fig:dependency_example}).

To identify the dependent perturbations, we develop a greedy-based method (see the detailed algorithm on the website~\cite{ourwebsite}) to check for dependencies between current operator $o$ and the existing sub-sequence $Seq$. If $o$ shares dependencies with any operators in $Seq$, it is added to that group; otherwise, $o$ is placed into a new sub-sequence, indicating no dependencies with existing groups. Specifically, dependency checking involves a use-def analysis~\cite{tok2006efficient} where the target nodes $V'$ and edges $E'$ created by an operator $o'$ (the definition) are examined against the usage in operation $o$. 
If $o$ utilizes any nodes or edges defined by $o'$, a dependency exists. 
Taken the example in the bottom part of Figure~\ref{fig:dependency_example}, $AddNode(i)$ defines node $i$, and $RemoveNode(i)$ uses node $i$, establishing a dependency that necessitates grouping these operations together as an atomic operation to ensure safe crossover and mutation.

%这段话的作用是什么
% Note that the initial sequence ($o_1, o_2, \ldots, o_n$) 
% can be guaranteed to be valid through an on-the-fly check. 
% Specifically, starting with the FCG, a valid operation is randomly selected that can feasibly be applied to the current state of the FCG. After applying the operation, the FCG is updated accordingly. Subsequent operations are then chosen based on this updated FCG, ensuring each selected operation remains feasible.
{Note that, during the GA initialization, the initial sequence ($o_1, o_2, \ldots, o_n$) is guaranteed to be valid through the on-the-fly check. 
Specifically, starting with the initial graph $G$, a valid sub-sequence is randomly selected that can feasibly be applied to the current state of $G$. After applying this sub-sequence, the $G$ is updated to $G'$. Subsequent operators are then chosen based on this updated $G'$, ensuring each selected operator remains feasible.}

% The FCG is then updated, and subsequent operations are chosen based on this updated state, ensuring feasibility throughout.
% Subsequent operations are then chosen based on this updated FCG, ensuring each selected operation remains feasible.

During GA iterations, the generation of individuals differs significantly from this initial process. Instead of being generated on-the-fly, the entire sequences in individuals are first constructed (with crossover and mutation) and then evaluated later. It introduces the problem in maintaining the feasibility of each operator within the sequence. Our dependency analysis is designed to mitigate this challenge in crossover and mutation.

% \vspace{-0.1cm}
\subsection{Step 2.2 and 2.3: Crossover and Mutation}\label{methodology:crossover_mutation}
% \vspace{-0.1cm}
Once dependency-aware individuals are established, crossover and mutation processes are conducted at the sub-sequence level. This approach is crucial for maintaining the integrity of dependent operators within each individual.

\textbf{Crossover.}
Sub-sequences that contain dependent operators are randomly selected either to be retained or removed in their entirety from the new individual during the crossover process. This method helps prevent conflicts that could arise from breaking apart interdependent operations (see step 2.2 in Figure~\ref{fig:overview}).

\textbf{Mutation.}
% Mutation aims to enhance the diversity of the GA's population. 
As shown in step 2.3 of Figure~\ref{fig:overview}, a sub-sequence is randomly chosen, and one of three types of mutations is applied: \textit{adding}, \textit{removing}, or \textit{updating}.
{\textit{Adding} involves inserting a new operator at a randomly selected position within a random sub-sequence, while \textit{removing} deletes the operator at that position. 
\textit{Updating} involves replacing an existing operator within the sub-sequence with another random operator.} 

{Due to the possibility of disturbing the dependency of the sequence, we then perform an on-the-fly dependency check for operators.
If a mutation renders subsequent operators infeasible, such as by altering dependent nodes or edges, we use a fix strategy. Problematic operators may be modified to fit the new context (e.g., changing an edge or node) or removed from the sequence. If the fix fails, the mutation is abandoned.}

% If a mutation renders subsequent operators infeasible, for example, if it removes or alters nodes or edges on which these operators depend, we adopt a fix strategy. The problematic operators might be modified to adapt to the new context (e.g., changing one edge or node), or removed from the sequence. If a fix is not successful, the current mutation will be abandoned.

% \vspace{-0.25cm}
\subsection{Step 2.4: Evaluation and Selection}\label{methodology:fitness}
% \vspace{-0.15cm}
{After crossover and mutation in the GA, we obtain new individuals as offspring. The fitness function is crucial for selecting superior individuals from the offspring.}

The typical fitness function in adversarial attacks uses the model's output to decrease the prediction probability of the current class or increase that of the target class~\cite{ma2021graph, li2023black}. 
However, relying solely on model output may not be effective in the context of AMD attacks, especially due to the non-differentiability of instance-based models and decision trees. Specifically, it can lead to premature convergence (i.e., all yield similar probability values), particularly if the model consistently exhibits high confidence in classifying certain samples as malicious. 

% \vspace{-0.25cm}
\subsubsection{Fitness for MLP Model}\label{fitness_mlp}
To overcome this challenge, we introduce an additional guidance mechanism based on feature interpretation, i.e., SHAP~\cite{SHAP2017}, a popular technique for understanding feature importance. 
% This approach informs and directs the optimization~\cite{fan2020can, liu2022explainable, soi2024enhancing, severi2021explanation}, 
% as SHAP values indicate the importance of each feature in a model's prediction.
Features with positive SHAP values positively contribute to the prediction, whereas negative SHAP values indicate a negative contribution.

When the GA encounters local optima without observable changes in model output, SHAP values allow us to monitor feature-level changes, offering a finer-grained criterion for selection. Specifically, \textit{if an individual increases the value of features with negative contributions or decreases the value of features with positive contributions, the prediction is closer to failure}, even if the probability output remains unchanged.

We define a multi-objective fitness function as follows:
% \vspace{-0.22cm}
\begin{equation}
\footnotesize
% \hspace{-2mm}
\begin{aligned}
    fitness1(I) = M(E(G+I)) \\
    fitness2(I) = - \sum_{i=0}^{n-1} SHAP(M, G, I)_i \cdot (E(G)_i - E(G+I)_i)
\end{aligned}
% \vspace{-0.18cm}
\end{equation}
where $I$ is a given individual (i.e., perturbations), $G$ is the original FCG, $E(G)$ is the embedding vector of the $G$ with length $n$, $M(E(G+I))$ is the probability of the benign class and $SHAP(M, G, I)$ represents the SHAP values of features.

The $fitness1$ evaluates the model's target class probability. The $fitness2$ measures the potential for classification changes, with $SHAP(M, G, I)_i$ indicating the direction (positive or negative) and $E(G)_i - E(G+I)_i$ quantifying the change in the $i$-{th} feature value due to perturbation $I$.

\textbf{Dominance and Selection.} 
We define a dominance relation for selection based on the two scores. An individual $x$ is said to dominate another one $y$ if and only if:
% \vspace{-0.18cm}
\begin{equation}
\label{eq:dominance}
\footnotesize
\begin{aligned}
    fitness1(x) > fitness1(y) \vee \\
    fitness1(x) == fitness1(y) \wedge fitness2(x)>fitness2(y)
    \end{aligned}
    % \vspace{-0.18cm}
\end{equation}
\vspace{0.1cm}
We prioritize individuals with higher benign probability scores or, when scores are equal, those that modify feature values in the most beneficial direction.

% \vspace{-0.25cm}
\subsubsection{Fitness for Instance-based Model}
For instance-based learning models (e.g. KNN), the main challenge is the lack of gradient information. To approximate gradients for KNN, we employ a surrogate model (e.g., an MLP model), which facilitates the use of an interpretation-based approach (see \S \ref{fitness_mlp}) alongside the model output.

For a KNN model $M$, where the adversarial challenge is to manipulate the instance such that it resembles benign samples more closely than malware samples, we train a surrogate model $M'$. This allows us to derive a dual-score fitness function, as follows:
% \vspace{-0.25cm}
\begin{equation}
\footnotesize
% \hspace{-2.7mm}
% \vspace{-0.15cm}
\begin{aligned}
    fitness1(I) = \frac{1}{x} \sum_{i=1}^{k} (M(I)^m_i - M(I)^b_i) \\
    fitness2(I) = - \sum_{i=0}^{n-1} SHAP(M', G, I)_i \cdot (E(G)_i - E(G+I)_i)
    \end{aligned}
\end{equation}
where $k$ represents the number of neighbors considered in KNN. $M(I)^m_i$  denotes the distance to the $i$-th nearest malware sample, and $M(I)^b_i$ denotes the distance to the $i$-th nearest benign sample. 

The $fitness1$ aims to increase the similarity to benign neighbors and decrease the similarity to malware neighbors, effectively manipulating the prediction of the adversarial example. The second fitness function $fitness2$, similar to that used for target MLP models (\S ~\ref{fitness_mlp}), utilizes SHAP values estimated by the surrogate model $M'$ to assess the impact of perturbations on feature importance. The selection follows the dominance relation defined in Equation~\ref{eq:dominance}.

% \vspace{-0.2cm}
\subsubsection{Fitness for Ensemble Model}

% Ensemble models (e.g., Random Forest) determine the model output probability through a voting process among numerous decision trees. Each tree in the model selects a subset of features to use as decision nodes (i.e., tree split nodes~\cite{song2015decision}). This selection makes altering the output values of tree models during adversarial attacks particularly challenging, especially when only a few or none of the selected decision features are present in the target sample. In such cases, it is unlikely to change the overall model output (i.e., the decision of the majority of trees) as the key decision features may remain unaffected.
Ensemble models like Random Forest determine output probabilities through a voting process among numerous decision trees, each selecting a subset of features for decision nodes (tree split nodes)~\cite{song2015decision}. Altering tree-based model outputs during testing is challenging, especially when only a few or none of the selected decision features are present in the target sample. Consequently, changes in the overall model output (i.e., the decision of the majority of trees) are unlikely if key decision features remain unaffected.

To overcome this challenge, we directly examine the constraints associated with the decision features. By analyzing the decision paths of all decision trees, we identify all possible feature constraints that could result in a benign output, as our goal is to have the target model misclassify the malware as benign. We will eliminate the constraints that conflict with those from other decision trees. Finally, our objective is to maximize the number of constraints that the perturbed inputs can satisfy, thereby increasing the likelihood of a benign classification. The fitness function is defined as follows:
% \vspace{-0.15cm}
\begin{equation}
    \footnotesize
   fitness(I) = \sum_{c\in C} SAT(G, M, I, c)
   % \vspace{-0.15cm}
\end{equation}
where $C$ represents all the constraints that can potentially lead to a benign output, and $SAT$ determines whether a given constraint $c\in C$ is satisfied (1) or not (0). The optimization process aims to generate perturbations that maximize the number of satisfied constraints.

%% file: tex/6-evaluation.tex
\section{Evaluation}\label{sec:evaluation}
% \vspace{-0.1cm}
We aim to evaluate the effectiveness of \tool by answering the following research questions.

\begin{itemize}[label=-,leftmargin=*]
    \item \textbf{RQ1}: How effective is \tool compared to others?
    \item \textbf{RQ2}: What is the performance of \tool?
    \item \textbf{RQ3}: How does each component of \tool impact the overall effectiveness?
    % \item \textbf{RQ4}: Can \tool adapt to black-box scenarios?
\end{itemize}

% \vspace{-0.2cm}
\subsection{Experimental Setup}\label{evaluation:setup}
% \vspace{-0.1cm}
\noindent\textbf{Dataset.}
Since the datasets used in previous studies are not publicly available, we adhere to the very common methodologies described in prior works~\cite{zhao2021structural,li2023black} to collect datasets. 
{The collected dataset includes 12,000 samples with 6,000 benign and 6,000 malware samples, divided into an 80:20 ratio for training and testing the models. To ensure representativeness, these collected samples are evenly distributed across six years, from 2018 to 2023, with 1,000 benign and 1,000 malware samples per year.} Similar to previous works~\cite{zhao2021structural, li2023black}, benign samples are sourced from AndroZoo~\cite{androzoo} (VirusTotal~\cite{virustotal} score of 0) and malware samples from VirusShare~\cite{virusShare} (VirusTotal score above 4).
To assess robustness, we additionally collected 120 true malware samples from the same six-year period (20 samples per year) as test seeds. 
{These seed samples are distinct from the initial set of 6,000 malware samples. More details about dataset are available on our website~\cite{ourwebsite}.}

\noindent\textbf{Target Models.} 
We invested significant efforts to include a wide range of models and features, ensuring a systematic and comprehensive evaluation of the testing methods. Specifically, we trained 40 (8×5) ML-based AMD models built upon 8 types of features, including the Degree, Katz, Harmonic, Closeness, Average, and Concentrate features from \malscan~\cite{wu2019malscan}, the family level from \mamadroid~\cite{onwuzurike2019mamadroid}, and the package level from \apigraph, alongside 5 ML-based classifiers: MLP~\cite{grosse2017adversarial}, KNN-1~\cite{fix1952discriminatory}, KNN-3, Random Forest (RF)~\cite{breiman2001random}, and AdaBoost (AB)~\cite{dai2018adversarial}. The performance of target models can be found on our website~\cite{ourwebsite}.
% We ensured the quality of target models, i.e., they achieve an average accuracy exceeding 90\%. The model details can be found on our website~\cite{ourwebsite}.

\noindent\textbf{Baselines.}
% \cs{here}
We selected three baselines for comparison: a random testing approach and two state-of-the-art adversarial attack methods~\cite{zhao2021structural,li2023black} for AMD. Due to the limited applicability of these state-of-the-art baselines or the unavailability of the code, we extended or re-implemented them based on the descriptions provided in their respective papers.

% Specifically, for \bagammo~\cite{li2023black}, which has not released its code, we replicated its algorithm based on the descriptions provided in their paper. 
Specifically, for \bagammo~\cite{li2023black}, which has not released its code, we replicated its algorithm based on the descriptions provided in their paper. 
{
To ensure a fair comparison with \tool, we configured \bagammo in a white-box setting, where feedback is obtained directly from the target model alongside a surrogate model. In the configuration referred to as \textit{BagAmmo-G}, we used a GCN surrogate model (the original model in this baseline) trained on ground truth data. Additionally, we experimented with using an MLP (the same surrogate model as in our method) in place of their original GCN, designated as \textit{BagAmmo-M}.
}
For \hrat~\cite{zhao2021structural}, although the code is available~\cite{hratcode}, it primarily addresses attacks utilizing Degree and Katz centrality metrics for \malscan on the KNN-1 model. We expanded its application to encompass a wider array of target scenarios. However, \hrat is limited by GPU memory constraints and the need for classifier differentiability~\cite{alzantot2019genattack, chen2020frank}, which restricts its use with tree-based models and memory-intensive features such as Average and Concentrate. 
{Regarding the Random attack, it randomly generates perturbation operators and evaluates their effects when applied to the FCG.}
Further details about the baselines, including the source code, are available on our website~\cite{ourwebsite} and GitHub~\cite{ourcode}.

% This method uses four basic perturbation operators, randomly generating 300 perturbation operators and applying them to the original FCG. The maximum number of iterations is 100.

% Since \bagammo~\cite{li2023black} has not disclosed the code, we reproduced a GA based on the descriptions in their paper. We used a five-layer GCN as a substitute model, where the GCN features include graph edges along with the nodes' in-degrees and out-degrees. 

% \textit{Random Attack.}

\noindent\textbf{Metrics.} We employed three widely used metrics: Attack Success Rates (ASR), Perturbation Rates (PR) and Average Number of Survival Genes per Generation (ASGG). ASR measures the effectiveness of attack methods. PR quantifies the relative increase in graph components (i.e., nodes and edges) of adversarial samples compared with the original malware. Considering the potential conflicts that can result in certain infeasible perturbations (i.e., genes in the individuals), ASGG is designed to assess the count of genes that remain feasible (referred to as the `survival genes') following crossover and mutation in each generation.
\vspace{-0.1cm}
\begin{equation}
\vspace{-0.2cm}
\footnotesize
    ASR = \frac{N_a}{N_m}, \
    PR = \frac{1}{N_a} \sum_{i=1}^{N_a} \delta_i, \
   ASGG= \frac{1}{G} \sum_{g=1}^{G} N_{g}
\end{equation}
where $N_a$ is the number of malware that can successfully bypass the AMDs, $N_m$ is the total number of seed malware,
$\delta_i = \frac{P_{add,i}}{P_{ori,i}}$ is the perturbation ratio for the $i$-th successful sample, reflecting the proportion of added nodes and edges,  $G$ is the total number of generations and $N_{g}$ is the total number of surviving genes in the $g$-{th} generation.

\begin{table*}[!t]
\centering
\caption{Attack Success Rates of \tool and Baselines.}
% \vspace{-0.3cm}
\label{tab:RQ1_asr}
\resizebox{.99\textwidth}{!}{
% \small

\begin{tabular}{c|ccccc|ccccc|ccccc|ccccc}
\hline
\multicolumn{1}{l|}{}                   & \multicolumn{5}{c|}{\malscan (Degree)}  & \multicolumn{5}{c|}{\malscan (Katz)}        & \multicolumn{5}{c|}{\malscan (Harmonic)} & \multicolumn{5}{c}{\malscan (Closeness)} \\ \cline{2-21} 
\multicolumn{1}{l|}{\multirow{-2}{*}{}} & MLP    & KNN-1  & KNN-3  & RF   & AB   & MLP     & KNN-1   & KNN-3  & RF    & AB    & MLP   & KNN-1  & KNN-3  & RF    & AB    & MLP   & KNN-1  & KNN-3  & RF    & AB    \\ \hline
\rowcolor[HTML]{CFE2F3} 
\textbf{Ours}                                    & \textbf{0.82}   & \textbf{0.73}   & \textbf{0.76}   & \textbf{0.78} & \textbf{1.00} & \textbf{0.77}    & \textbf{0.68}    & \textbf{0.69}   & \textbf{0.94}  & \textbf{0.96}  & \textbf{0.82}  & \textbf{0.90}   & \textbf{0.83}   & \textbf{1.00}  & \textbf{1.00}  & \textbf{0.88}  & \textbf{0.97}   & \textbf{0.84}   & \textbf{0.91}  & \textbf{1.00}  \\
\textbf{\hrat}                                     & 0.14   & 0.18   & 0.08   & -    & -    & 0.01    & 0.03    & 0.01   & -     & -     & 0.01     & 0.03      & 0.01      & -     & -     & 0.01     & 0.12      & 0.18      & -     & -     \\
\rowcolor[HTML]{CFE2F3} 
\textbf{\bagammo-M}                           & 0.58     & 0.57      & 0.50     & 0.02     & 0.07     & 0.24      & 0.23      & 0.16     & 0.01     & 0.03     & 0.70      & 0.66      & 0.61     & 0.37   & 0.09    & 0.77    & 0.67   & 0.59      & 0.13      & 0.13   \\

\textbf{\bagammo-G}                           & 0.65     & 0.61      & 0.28     & 0.03     & 0.06     & 0.43      & 0.22      & 0.05     & 0.00     & 0.03     & 0.73      & 0.67      & 0.58     & 0.33   & 0.08    & 0.74    & 0.64   & 0.56      & 0.05      & 0.08   \\
\rowcolor[HTML]{CFE2F3} 
\textbf{Random}                                  & 0.59   & 0.62   & 0.56   & 0.19 & 0.03 & 0.02    & 0.08    & 0.08   & 0.11  & 0.04  & 0.59  & 0.67   & 0.57   & 0.08  & 0.03  & 0.63  & 0.58   & 0.58   & 0.13  & 0.04  \\ \hline \hline 
\multicolumn{1}{l|}{}                   & \multicolumn{5}{c|}{\malscan (Average)} & \multicolumn{5}{c|}{\malscan (Concentrate)} & \multicolumn{5}{c|}{Mamadroid}          & \multicolumn{5}{c}{APIGraph}            \\ \cline{2-21} 
\multicolumn{1}{l|}{\multirow{-2}{*}{}} & MLP    & KNN-1  & KNN-3  & RF   & AB   & MLP     & KNN-1   & KNN-3  & RF    & AB    & MLP   & KNN-1  & KNN-3  & RF    & AB    & MLP   & KNN-1  & KNN-3  & RF    & AB    \\ \hline 
\rowcolor[HTML]{CFE2F3} 
\textbf{Ours}      &                              
% & 0.90   & 0.92   & 0.83   & 0.91 & 1.00 & 0.87    & 0.91    & 0.78   & 0.94  & 0.96  & 0.96  & 0.99   & 0.93   & 1.00  & 0.98  & 0.83  & 0.84   & 0.84   & 0.78  & 0.72  \\
\textbf{0.90} & \textbf{0.92} & \textbf{0.83} & \textbf{0.91} & \textbf{1.00} & \textbf{0.87} & \textbf{0.91} & \textbf{0.78} & \textbf{0.94} & \textbf{0.96} & \textbf{0.96} & \textbf{0.99} & \textbf{0.93} & \textbf{1.00} & \textbf{0.98} & \textbf{0.83} & \textbf{0.84} & \textbf{0.84} & \textbf{0.78} & \textbf{0.72} \\

\textbf{\hrat}                                     & -      & -      & -      & -    & -    & -       & -       & -      & -     & -     & 0.03     & 0.08      & 0.03      & -     & -     & 0.05     & 0.09      & 0.04      & -     & -     \\
\rowcolor[HTML]{CFE2F3} 
\textbf{\bagammo-M}                                 & 0.75      & 0.63      & 0.58      & 0.24    & 0.18    & 0.72       & 0.66       & 0.57      & 0.17     & 0.19     & 0.58  & 0.76   & 0.71   & 0.40  & 0.13  & 0.81  & 0.79   & 0.78   & 0.07  & 0.43  \\
\textbf{\bagammo-G}                                 & 0.72      & 0.62      & 0.57      & 0.05    & 0.11    & 0.73       & 0.60       & 0.55      & 0.23     & 0.06     & 0.80  & 0.65   & 0.53   & 0.33  & 0.33  & 0.79  & 0.74   & 0.70   & 0.16  & 0.40  \\
\rowcolor[HTML]{CFE2F3} 
\textbf{Random}                                  & 0.61   & 0.62   & 0.58   & 0.08 & 0.03 & 0.66    & 0.60    & 0.58   & 0.20  & 0.03  & 0.09  & 0.12   & 0.03   & 0.13  & 0.14  & 0.47  & 0.14   & 0.03   & 0.05  & 0.09  \\  \hline 
\end{tabular}
}
\vspace{-0.5cm}
\end{table*}

% \vspace{-0.1cm}
\subsection{RQ1: Effectiveness}\label{evaluation:RQ1 effectiveness}

% \vspace{-0.15cm}
\noindent\textbf{Setup.}
We initialize each population with 100 individuals for 40 generations, with each individual initializing with 300 perturbation operations. This configuration follows established precedents in the literature. According to \bagammo, the best attack success rate is achieved at 40 generations. Meanwhile, \hrat identifies 300 as the optimal number of perturbations. To enhance \hrat's performance and ensure fair comparisons, we increased the number of random initializations in \hrat from 16 to 100. The configuration for the Random attack remains consistent with the baselines previously described, involving 300 perturbations per iteration and a maximum number of 100 iterations.

\noindent\textbf{Results \& Analysis.}
As presented in Table~\ref{tab:RQ1_asr}, the results demonstrate that our attack method outperforms the baselines significantly, the results demonstrate that our attack method significantly outperforms the baselines, achieving an average ASR of 87.9\%, {which is at least 44.7\% higher than \bagammo-M (43.2\%), \bagammo-G (41.2\%), Random Attack (28.8\%), and \hrat (7.8\%).}

(1) Baseline Analysis: 
%We found that  \hrat generally performs poorly across most models, often yielding the ASR close to zero, especially in MalScan (Katz, Harmonic), \mamadroid, and \apigraph, where it is ineffective, which has a huge discrepency with their claimed in their paper, similar doubts in this survey~\cite{olszewski2023get} and its results\footnote{https://github.com/reproducibility-sec/reproducibility/blob/main/sheet1.csv}. 
We found that \hrat generally performs poorly across most models, often yielding an ASR close to zero, especially in MalScan (Katz, Harmonic), \mamadroid, and \apigraph, where it is ineffective. 
Furthermore, the results of \malscan (Degree and Katz) with KNN-1 show a significant discrepancy compared to the claims in their paper, with similar doubts raised in this survey~\cite{olszewski2023get} and its results.\footnote{https://github.com/reproducibility-sec/reproducibility/blob/main/sheet1.csv}
% We observed that during the model's forward propagation, the changes in feature values tend towards zero, leading to gradient vanishing, which is a well-known problem in training deep neural networks~\cite{liu2020loss, hameed2021gradient,liu2023enhancing}. 
% It is particularly challenging when the detector uses different feature methods or models (e.g., KNN).
% To address this, it is necessary to carefully adjust specific parameters (e.g., learning rate) or optimize configurations (e.g., activation function) 
% when switching between different feature methods or models, which requires a significant amount of human effort. 
% Our analysis found that the primary limitation of \hrat is that the reward mechanism in reinforcement learning relies on coarse model scores and the inaccuracy of gradient estimation on the adjacency matrix.
% Consequently, this method struggles to adapt to diverse target models; 
{Our analysis indicates that a primary limitation of \hrat lies in its RL reward mechanism, which relies on coarse model scores and suffers from inaccuracies in gradient estimation on the adjacency matrix. This lack of precise, granular feedback hinders the model’s ability to fine-tune perturbations across complex feature types and target models.
\bagammo-M and \bagammo-G perform well with \mamadroid and \apigraph feature types across MLP, KNN-1, and KNN-3 models, especially with MLP. However, their effectiveness drops significantly on tree-based models and \malscan, where they perform comparably to Random Attack. This reduction stems primarily from reliance on a single mutation operator and feedback based solely on coarse-grained scores from surrogate and target models. Although it claims GCN’s strengths with graph-structured data~\cite{li2023black}, adapting to tree-based models in AMD is challenging, as these models typically lack the relational structure that GCNs leverage. Please note that we observed a discrepancy between our evaluation results and those reported in the original paper. 
We have made extensive efforts to investigate this issue and confirmed our results, as discussed in Section~\ref{threatstovalidity} and the explanations on our website~\cite{ourwebsite}.}

% \textbf{Finding #1:}
% \textit{Existing methods are
% notably ineffective, particularly on MalScan and ensemble models, with results that are close to random attacks.}

\vspace{-0.15cm}
\begin{tcolorbox}[left=1pt, right=1pt, top=0pt, bottom=0pt]
\label{insight:1}
\small
% \hypertarget{}
{\textbf{Finding \#1}}: 
\textit{
Existing methods are notably ineffective, particularly on \malscan and ensemble models, with results that are close to random testing.
}
\end{tcolorbox}
% \vspace{-0.15cm}

% This decrease can be attributed to the fundamental design of GCNs, which are optimized for graph-structured data where node connections are crucial~\cite{sun2022adversarial}. In contrast, tree-based models in AMD often rely on features derived from static analysis that lack such relational structures. Furthermore, the extensive dimensionality of \malscan creates a larger search space compared to \mamadroid and \apigraph, which challenges the efficacy of \bagammo under these conditions.

% \vspace{-0.15cm}
% \begin{tcolorbox}[left=1pt, right=1pt, top=0pt, bottom=0pt]
% \label{insight:1}
% \small
% \hypertarget{}
% {\textbf{Finding \#1}}: 
% \textit{
% %强化学习不适用于广泛的机器学习检测器。基于GCN的黑盒攻击不能在esemble model上产生有效的攻击。
% Gradient-based methods are not suitable for non-differentiable models. Black-box attacks using GCNs and  single type of perturbation fail to produce effective attacks on ensemble models and \malscan.
% }
% \end{tcolorbox}
% \vspace{-0.1cm}

(2) Robustness Analysis:
There are noticeable variations in ASR across different feature types. 
From the perspective of features, \malscan (Closeness and Average) and \mamadroid consistently exhibit high ASRs, typically exceeding 90\% across most classifiers, suggesting these features may be more susceptible to attacks. Conversely, features like \malscan (Degree and Katz) and \apigraph demonstrate greater robustness, with ASRs often below 80\%, likely due to the complexity of perturbing these features effectively; While ensemble models (i.e., RF and AB) generally show more robustness than single models (i..e, MLP and KNNs), our method still achieves high ASRs on \malscan (Harmonic) and \malscan (Closeness) features, proving its effectiveness even against complex models. However, in ensemble models, our method achieves relatively lower ASRs, notably less than 80\% on \apigraph, reflecting their resilience against attacks.

% \textbf{Finding #2:}
% \textit{\malscan (Closeness and Average) and \mamadroid are attack-prone, while \malscan (Degree and Katz) and \apigraph are more robust; Ensemble models (i.e., RF and AB) generally show greater robustness.}

\vspace{-0.15cm}
\begin{tcolorbox}[left=1pt, right=1pt, top=0pt, bottom=0pt]
\label{insight:2}
\small
% \hypertarget{}
{\textbf{Finding \#2}}: 
\textit{
\malscan (Closeness and Average) and \mamadroid are less robust, while \malscan (Degree and Katz) and \apigraph are more robust; Ensemble models (i.e., RF and AB) generally show greater robustness.
}
\end{tcolorbox}

% \cs{Here, number}

% \tool significantly outperforms baseline approaches in white-box attack scenarios, achieving higher ASR across diverse target models. Furthermore, the results underscore the persistent susceptibility of current FCG-based ML/DL models to adversarial attacks, emphasizing the need for enhanced security testing.

\vspace{-0.25cm}
\begin{tcolorbox}[left=1pt, right=1pt, top=0pt, bottom=0pt]
\label{conclusion:1}
\small
% \hypertarget{}
{\textbf{Answer to RQ1}}: 
\textit{\tool, with an average ASR of 87.9\%, outperforms baselines by at least 44.7\% in white-box attacks, achieving higher ASR across diverse models.
The results highlight the persistent vulnerability of current FCG-based ML models to adversarial attacks, emphasizing the need for robustness testing.
% \tool outperforms baselines in white-box attacks, achieving higher ASR across diverse models.
% The results highlight the persistent vulnerability of current FCG-based ML models to adversarial attacks, emphasizing the need for robustness testing.
}
\end{tcolorbox}
\vspace{-0.5cm}
%%%%%%%%%%%%%%%%%%%%%%%%%%%%%%%%%%%%%%%%%%%%%%%%
\begin{table}[!t]
\centering
\caption{Perturbation Rates of Successful Attack Samples.}
% \vspace{-0.2cm}
\label{tab:rq2_perturbation}
\resizebox{.48\textwidth}{!}{
\begin{tabular}{c|c|c|c|c|c}
\hline
\multicolumn{1}{l|}{}  & MLP            & KNN-1          & KNN-3          & RF               & AB               \\ \hline
\rowcolor[HTML]{CFE2F3} 
\textbf{\malscan (Degree)}      & 0.04           & 2.23           & 0.04           & 5.17             & 2.29             \\
\textbf{\malscan (Katz)}        & 0.17           & 2.88           & 0.19           & 4.68             & \textgreater{}10 \\
\rowcolor[HTML]{CFE2F3} 
\textbf{\malscan (Harmonic)}    & \textbf{\textless 0.01} & 0.81           & \textbf{\textless 0.01} & \textbf{0.02}             & \textbf{0.01}             \\
\textbf{\malscan (Closeness)}   & \textbf{\textless 0.01} & 0.52           & \textbf{\textless 0.01} & 0.68             & 1.58             \\
\rowcolor[HTML]{CFE2F3} 
\textbf{\malscan (Average)}     & \textbf{\textless 0.01} & \textbf{\textless 0.01} & \textbf{\textless 0.01} & 5.27             & 8.07             \\
\textbf{\malscan (Concentrate)} & \textbf{\textless 0.01} & 1.70           & \textbf{\textless 0.01} & \textgreater{}10 & \textgreater{}10 \\
\rowcolor[HTML]{CFE2F3} 
\textbf{Mamadroid}             & 0.45           & 0.45           & 0.14           & \textgreater{}10 & 4.85             \\
\textbf{APIGraph}              & 1.31           & 0.26           & 0.14           & \textgreater{}10 & 3.07             \\ \hline
\end{tabular}
}
\vspace{-0.15cm}
\end{table}

% \vspace{0.1cm}
\subsection{RQ2: Performance}\label{evaluation:RQ2 performance}
% \vspace{-0.25cm}
\subsubsection{Perturbation Rates across Target Models}\label{evaluation:RQ2 performance:perturbation}
\hfill \\
\textbf{Setup.}
To measure the perturbation degree that \tool applies to original samples, we calculated the average PR across all adversarial samples for each target model.

\noindent\textbf{Results \& Analysis.}
As shown in Table \ref{tab:rq2_perturbation}, the successful adversarial samples on MLP and KNNs-based models exhibit relatively low PR, while higher PR on ensemble models. 
Specifically, \malscan (Harmonic, Closeness, Average, and Concentrate) models achieved PRs below 0.01 with both MLP and KNN-3. 
However, PRs for KNNs are slightly higher compared to MLP, suggesting MLP's less robustness. PRs for tree-based models (i.e., RF and AB) are notably higher, except for \malscan (Harmonic), where they are lower. 
Instances of PRs exceeding 10 are observed, typically due to exceptionally large samples. For instance, in \malscan (Katz) under AB, 50\% of the samples are below 3, and 73\% are below 10. 

% Overall, higher PRs indicate that ensemble models are more challenging to attack than KNNs and MLP models, yet the scale of applied perturbations in successful attacks remains within a reasonably acceptable range.

% \vspace{-0.15cm}
\begin{tcolorbox}[left=1pt, right=1pt, top=0pt, bottom=0pt]
\label{insight:3}
\small
% \hypertarget{}
{\textbf{Finding \#3}}: 
\textit{Single models (i.e., MLP and KNNs) are generally easier to bypass, requiring fewer perturbations. In contrast, ensemble models (i.e., RF and AB) combine predictions from multiple models, making them more robust and necessitating greater perturbations to compromise.
}
\end{tcolorbox}
% \vspace{-0.15cm}

% \vspace{-0.05cm}
\subsubsection{Survival Genes during GA Iterations}\label{evaluation:RQ2 performance:survival_genes}
\hfill \\
\textbf{Setup.}
To assess the usefulness of the dependency-aware strategy (see \S ~\ref{methodology:individual_representation}) in reducing mutation conflicts, we calculated the ASGG after 40 iterations. Specifically, we randomly selected 20 malware and conducted experiments with and without the dependency analysis for comparative analysis. 
% To prevent premature termganination of these samples, the evaluation phase of the GA was omitted. 

% \vspace{-0.55cm}

\noindent\textbf{Results \& Analysis.}
The green and orange lines in Figure~\ref{fig:rq2_survival_genes} represent \tool's ASGG with and without the dependency-aware strategy, respectively. Throughout the iterations, the green line consistently maintains a higher ASGG compared to the orange line, with the difference doubling after the fifth generation. The orange line shows a noticeable bump between generations 30 and 35. Our analysis indicates that this increase can be attributed to the probabilistic introduction of a significant number of new genes by the mutation, leading to pronounced fluctuations.
% As detailed in \S~\ref{methodology:crossover_mutation}, our analysis shows that this increase can be attributed to the mutation process, which probabilistically introduces a significant number of new genes, potentially leading to pronounced fluctuations. 
Following this bump, the ASGG quickly declines due to the absence of the dependency-aware strategy capable of preemptively resolving gene conflicts. In contrast, the green line remains more stable throughout the generations. This stability suggests that the strategy helps preserve the number of viable genes within the population, which could prevent premature convergence of the GA optimization. Stability in the gene pool is crucial because significant diminishment in genetic variety can impede the GA’s ability to generate new and potentially more effective individuals~\cite{squillero2016divergence}.

% \vspace{-0.15cm}
\begin{tcolorbox}[left=1pt, right=1pt, top=0pt, bottom=0pt]
\label{insight:4}
\small
% \hypertarget{}
{\textbf{Finding \#4}}: 
\textit
{
The dependency-aware strategy protects critical genes from mutation conflicts, thereby preserving genetic diversity and preventing premature convergence by maintaining a sufficient number of viable perturbations.
}
\end{tcolorbox}
% \vspace{-0.15cm}

% \vspace{-0.05cm}
\subsubsection{Runtime Performance Compared with Others}\label{evaluation:RQ2 performance:runtime}
\hfill \\
\textbf{Setup.}
% To evaluate the attack costs, we tracked the number of adversarial samples generated within 500 minutes on \malscan (Degree) and KNN-1 as the target model where \hrat has the highest ASR.
To assess the performance of the testing, we monitored the number of adversarial samples generated within 500 minutes on \malscan (Degree) and KNN-1 models. These models and features were chosen because the baselines (e.g., \hrat) achieved the highest ASR with them.

\begin{figure}[!t]
\vspace{-0.2cm}
  \centering
\captionsetup{justification=centering}
% \hspace{0.15\textwidth} % 左边距
  \begin{minipage}[b]{0.24\textwidth}
    \includegraphics[width=\textwidth]{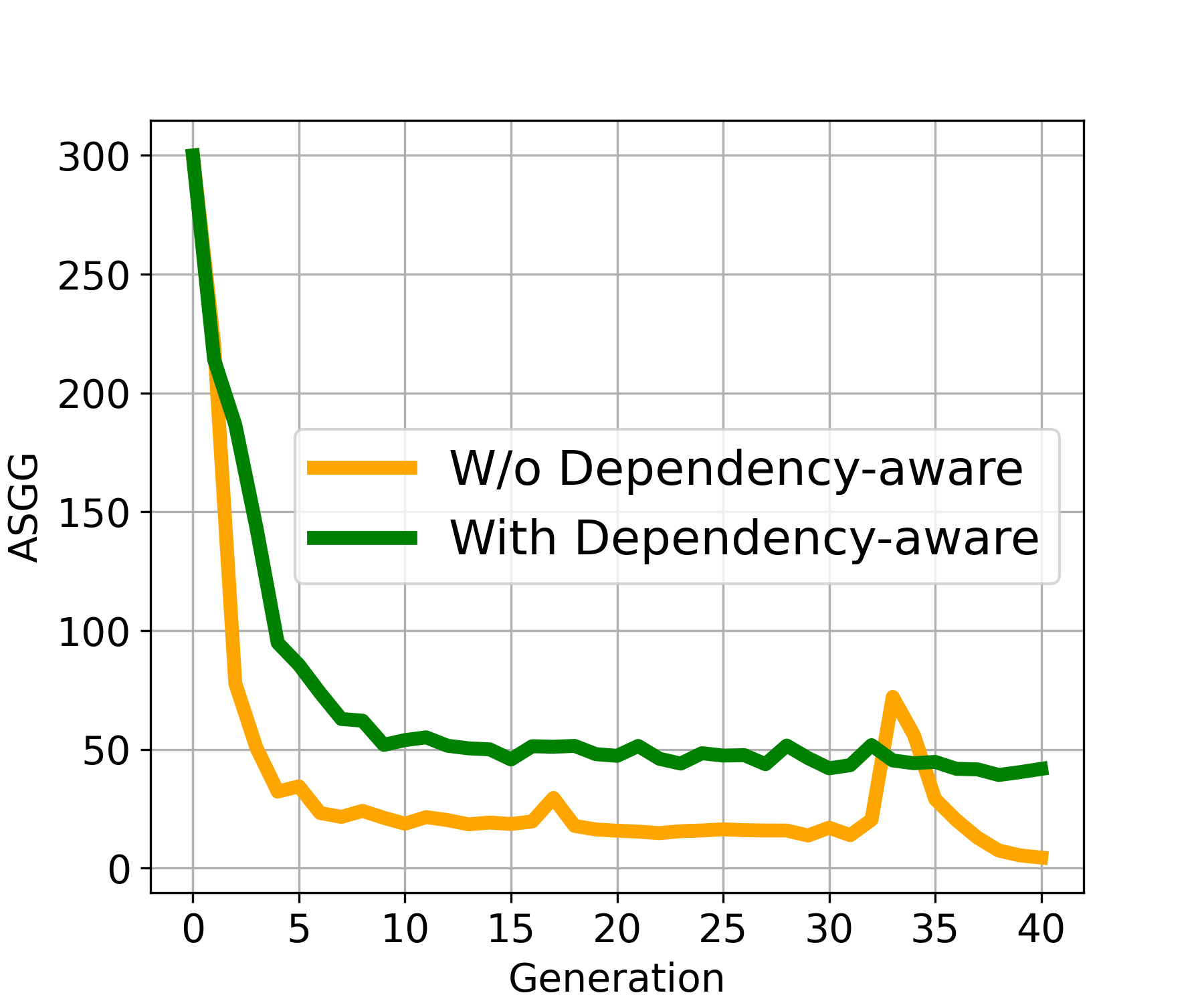}
    \caption{Impact of \\ Dependency-aware \\Strategy.}
    \label{fig:rq2_survival_genes}
  \end{minipage}
  \hfill
  \begin{minipage}[b]{0.24\textwidth}
    \includegraphics[width=\textwidth]{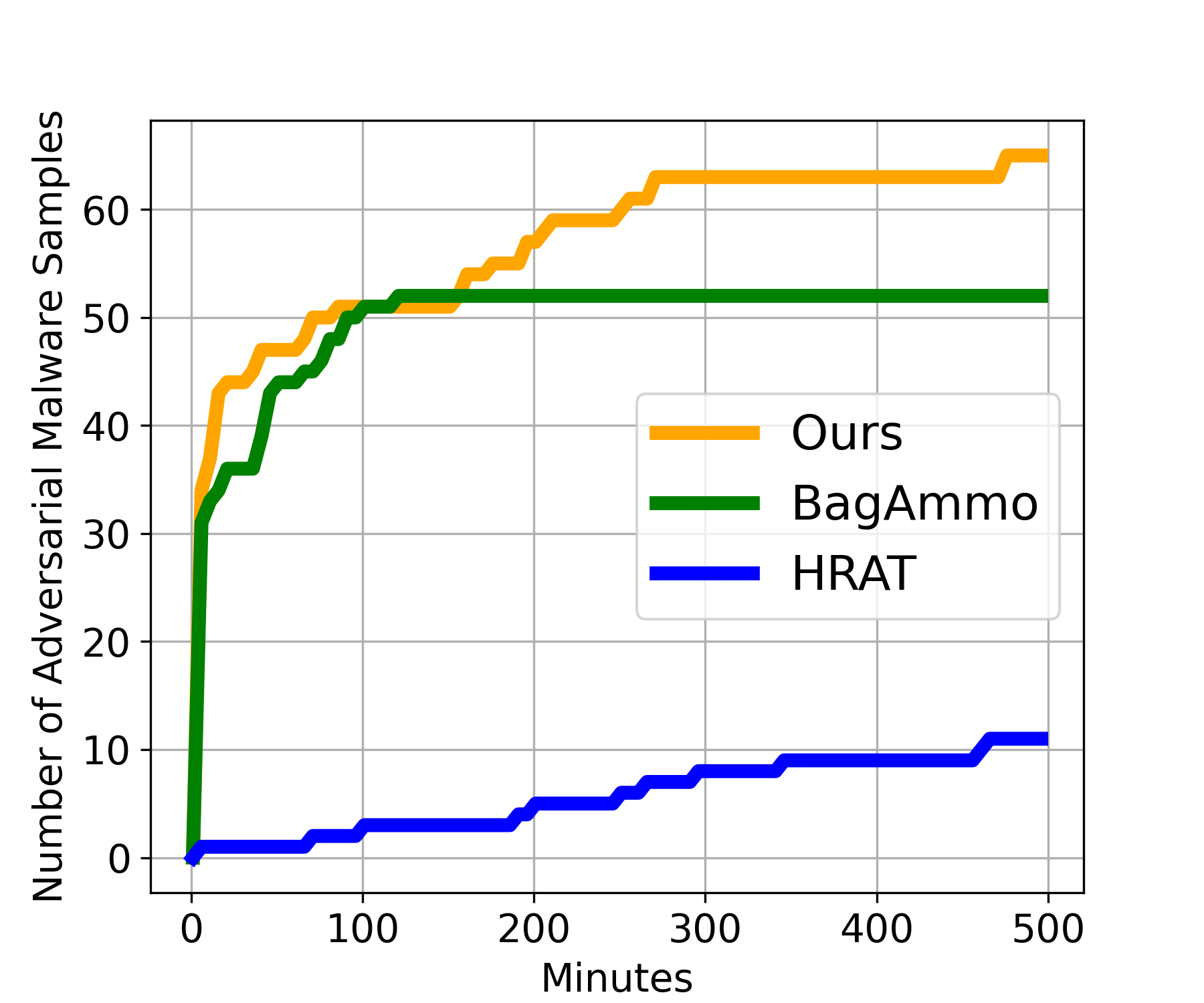}
    \caption{Runtime Efficiency Comparison with Other State-of-the-Art Methods.}
    \label{fig:runtime}
  \end{minipage}
% \hspace{0.15\textwidth} % 右边距

  % \vspace{-0.25cm}
\end{figure}

\noindent\textbf{Results \& Analysis.}
Figure~\ref{fig:runtime} demonstrates that \tool (orange line) detects more successful attacks than both \bagammo (green line) (i.e., \bagammo-G) and \hrat (blue line). 
\hrat's runtime efficiency is notably low because each perturbation requires the computation of gradient information from the FCG and target model. Although \tool and \bagammo efficiently completed most attacks within a short period, exhibiting exceptional runtime efficiency, \bagammo's progress stalls after approximately 110 minutes, indicating premature convergence, likely due to the limited variety in its single operators (i.e., only involve adding edges) and the lack of directional feedback from target models. In contrast, the persistent growth of \tool highlights its ability to continuously explore the huge search space and identify viable solutions.

{For further analysis, we also calculated the average time cost per iteration of mutant generation, with the following results: 0.34s for \bagammo, 18.72s for \hrat, and 0.67s for \tool. \hrat takes significantly longer due to the heavy gradient calculation. Compared to single-objective \bagammo, the selection process in our method is slightly slower. However, our dependency-aware mutation and multi-objective optimization are well worth it, as they ultimately lead to a significant reduction in the overall time cost required to detect adversarial exampless, as shown in Figure~\ref{fig:runtime}.}

% \vspace{-0.15cm}
\begin{tcolorbox}[left=1pt, right=1pt, top=0pt, bottom=0pt]
\label{conclusion:2}
\small
% \hypertarget{}
{\textbf{Answer to RQ2}}: 
\textit{
\tool is efficient in generating failure cases (i.e., more failures within the same time), adding fewer perturbations (i.e., lower PR), and maintaining a higher number of valid perturbations (i.e., higher ASGG).
% Successful attacks exhibited relatively low perturbation rates, indicating high attack efficiency. Notable efficiency was demonstrated in solving operator conflicts, thereby ensuring \ssw{diversified sequences to avoid premature convergence. \tool has also better runtime compared to other methods.}
}
\end{tcolorbox}
\vspace{0.35cm}

%%%%%%%%%%%%%%%%%%%%%%%%%%%%%%%%%%%%%%%%%%%%%%%%
\begin{table*}[!t]
\centering
% \caption{Results of Ablation Studies on the Key Components of \tool. 
\caption{Ablation Studies on the Key Components of \tool. 
}
% \vspace{-0.35cm}
\label{tab:rq3_ablation}
\resizebox{.99\textwidth}{!}{
\begin{tabular}{c|ccc|cccc|cccc|ccc|ccc}
\hline
                      & \multicolumn{3}{c|}{MLP} & \multicolumn{4}{c|}{KNN-1}    & \multicolumn{4}{c|}{KNN-3}    & \multicolumn{3}{c|}{Random Forest} & \multicolumn{3}{c}{AdaBoost} \\ \cline{2-18} 
\multirow{-2}{*}{}    & -Cri   & -Dep   & -Int   & -Cri  & -Dep  & -Int  & -Sur  & -Cri  & -Dep  & -Int  & -Sur  & -ASN       & -ADN      & -ALE      & -ASN     & -ADN    & -ALE    \\ \hline
\rowcolor[HTML]{CFE2F3} 
\malscan (Degree)      & -0.11  & \ul{-0.08}  & -0.12  & -0.12 & \ul{-0.03} & -0.05 & \ul{-0.03} & -0.15 & \ul{-0.09} & -0.14 & -0.16 & -0.01      & \ul{0.00}      & \ul{-0.19}     & \textbf{-0.19}    & 0.02    & -0.56   \\
\malscan (Katz)        & \textbf{-0.29}  & \textbf{-0.60}  & \textbf{-0.56}  & \textbf{-0.23} & -0.17 & -0.11 & -0.11 & \textbf{-0.23} & \textbf{-0.50} & -0.12 & -0.19 & -0.05       & -0.13     & -0.38     & -0.01    & -0.28   & -0.48   \\
\rowcolor[HTML]{CFE2F3} 
\malscan (Harmonic)    & -0.10  & \ul{-0.08}  & -0.07  & -0.18 & -0.17 & -0.18 & -0.13 & -0.18 & -0.13 & -0.15 & \textbf{-0.25} & -0.01       & \ul{0.00}      & -0.41     & \ul{0.00}     & -0.01    & -0.60   \\
\malscan (Closeness)   & -0.16  & -0.21  & -0.08  & -0.18 & -0.18 & -0.16 & -0.21 & -0.17 & -0.13 & -0.13 & -0.15 & -0.01       & -0.02     & -0.35     & -0.06    & -0.03    & -0.25   \\
\rowcolor[HTML]{CFE2F3} 
\malscan (Average)     & -0.20  & -0.15  & -0.12  & -0.18 & -0.18 & -0.18 & \textbf{-0.28} & -0.20 & -0.18 & \textbf{-0.18} & -0.21 & -0.03      & -0.02     & -0.23     & \ul{0.00}     & -0.13    & \ul{-0.18}   \\
\malscan (Concentrate) & -0.18  & -0.15  & -0.11  & \textbf{-0.23} & \textbf{-0.27} & \textbf{-0.23} & \textbf{-0.28} & -0.14 & -0.11 & -0.09 & -0.16 & \textbf{-0.06}       & \textbf{-0.14}     & -0.61     & -0.05    & \textbf{-0.50}   & -0.58   \\
\rowcolor[HTML]{CFE2F3} 
\mamadroid             & -0.18  & -0.20  & -0.16  & -0.18 & -0.22 & -0.19 & -0.20 & -0.14 & -0.22 & -0.18 & -0.16 & \ul{0.00}       & \ul{0.00}      & \textbf{-0.75}     & \ul{0.00}     & \ul{0.00}    & \textbf{-0.78}   \\
\apigraph              & \ul{-0.04}  & -0.13  & \ul{-0.04}  & \ul{-0.05} & -0.05 & \ul{-0.03} & \ul{-0.03} & \ul{-0.01} & -0.18 & \ul{-0.03} & \ul{-0.04} & \ul{0.00}       & \ul{0.00}     & -0.68     & \ul{0.00}     & \ul{0.00}    & -0.72   \\ \hline
\end{tabular}
}
\vspace{-0.25cm}
\end{table*}

\vspace{-0.55cm}

% \vspace{0.15cm}
\subsection{RQ3: Ablation Studies}\label{evaluation:RQ3 ablation}
% \vspace{-0.15cm}
\noindent\textbf{Setup.}
To assess the effectiveness of key components in \tool, we quantified the reduction in ASR across 40 target models upon the removal of the specific component. Specifically, key components include critical area identification (\S~\ref{methodology:critical_area_identification}) as $Cri$, dependency-aware strategy (\S~\ref{methodology:individual_representation}) as $Dep$, and fitness function (\S~\ref{methodology:fitness}). Additionally, three new perturbation operators (\S~\ref{sec:perturbation}), including Add Sparse Nodes ($ASN$), Add Dense Nodes ($ADN$), and Add Long Edges ($ALE$), are used mainly for ensemble models (i.e., RF and AB). 
% This is because these models exhibit very robust performance with low ASR (\S~\ref{evaluation:RQ1 effectiveness}) in random attack and higher PR (\S~\ref{evaluation:RQ2 performance:perturbation}).
The fitness function for MLP models incorporates interpretation-based scores ($Int$), whereas, for KNN models, it includes additional scores derived from a surrogate model($Sur$). To summarize, the evaluated components for MLP and KNNs include $Cri$, $Dep$, and $Int$, with $Sur$ additionally assessed for KNNs. For ensemble models, we focus on the impact of $ASN$, $ADN$, and $ALE$. The experimental parameters are consistent with those in RQ1 (\S~\ref{evaluation:RQ1 effectiveness}).

\noindent\textbf{Results \& Analysis.}\label{MLP_ablation}
Table~\ref{tab:rq3_ablation} displays the ASR discrepancies resulting from the removal of individual components, compared to the complete configuration in Table~\ref{tab:RQ1_asr}.

First, the removal of certain components results in notable reductions in ASR, emphasizing their critical role in the effectiveness of \tool. For instance, removing $Cri$ and $Dep$ significantly affects the ASR in MalScan (Katz) models more than in other models. The removal of $Cri$ always results in the largest drops, and under the KNN-1 and KNN-3 configurations, the removal of $Dep$ leads to decreases as high as 0.60 and 0.50, respectively. 
Due to its inherent robustness, MalScan (Katz) requires substantial perturbations (Table \ref{tab:rq2_perturbation}) and effective dependency management to maintain a viable number of genes in the population (Figure.~\ref{fig:rq2_survival_genes}), which prevents premature convergence of the GA, as noted in \S~\ref{evaluation:RQ2 performance:survival_genes}.
Moreover, the $Sur$ is critical for KNN models, where its removal leads to significant ASR reductions to other components. This suggests that KNN classifiers may rely more heavily on specific interpretations or feature relations that the surrogate model helps to exploit.

Second, certain features and models demonstrate minimal impact from the removal of components. 
% For instance, \apigraph shows notably smaller declines in ASR in configurations involving the removal of $Cri$, $Int$, and $Sur$, all below 0.04.
For instance, \apigraph shows notably smaller ASR declines, all below 0.04, when $Cri$, $Int$, and $Sur$ are removed.
This suggests that \apigraph has inherent robustness, as noted in \S~\ref{evaluation:RQ1 effectiveness}. 
Its robustness can be attributed to high-level feature abstraction, making it less sensitive to perturbations affecting fewer nodes or less critical connections within the FCG's structure.

Third, the impact of new perturbation operators varies significantly across ensemble models. The removal of $ALE$ demonstrates substantial ASR drops, especially in the \mamadroid and \apigraph models. This is primarily because these models are based on features constructed from the call relationships between functions, making them highly sensitive to substantial changes in edges, particularly those directed toward system functions. Although the node-adding based operators, $ASN$ and $ADN$, aim to reduce the centrality of malicious nodes in \malscan graphs, $ALE$ more significantly disrupts node centrality by altering graph structures, i.e, adding edges that modify path lengths and node connectivity (\S \ref{sec:perturbation}).
This change impacts the graph's overall centrality more drastically than simply adjusting nodes.

% \vspace{-0.15cm}
\begin{tcolorbox}[left=1pt, right=1pt, top=0pt, bottom=0pt]
\label{insight:5}
\small
% \hypertarget{}
{\textbf{Finding \#5}}: 
\textit{
%在最鲁棒的模型中，基于边的扰动比基于点的扰动更具有影响力。
Edge-based perturbations tend to be more effective than node-based ones, impacting the most robust models by altering the graph's features significantly.
}
\end{tcolorbox}

% \vspace{-0.25cm}
\begin{tcolorbox}[left=1pt, right=1pt, top=0pt, bottom=0pt]
\label{conclusion:3}
\small
% \hypertarget{}
{\textbf{Answer to RQ3}}:
\textit{
Each component in \tool plays a distinct role in enhancing the ASR for different models. Specifically, $Cri$ and $Dep$ significantly boost ASR in \malscan (Katz), while $Sur$ is crucial for KNN models. In ensemble models, $ALE$ proves to be highly influential.
}
\end{tcolorbox}
\vspace{-0.25cm}

%% file: tex/8-limitations_and_discussion.tex
\section{Discussions}
% \vspace{-0.15cm}
% \begin{enumerate}[leftmargin=*]
%     \item 

\begin{figure}
  \centering
  \includegraphics[width=0.48\textwidth]{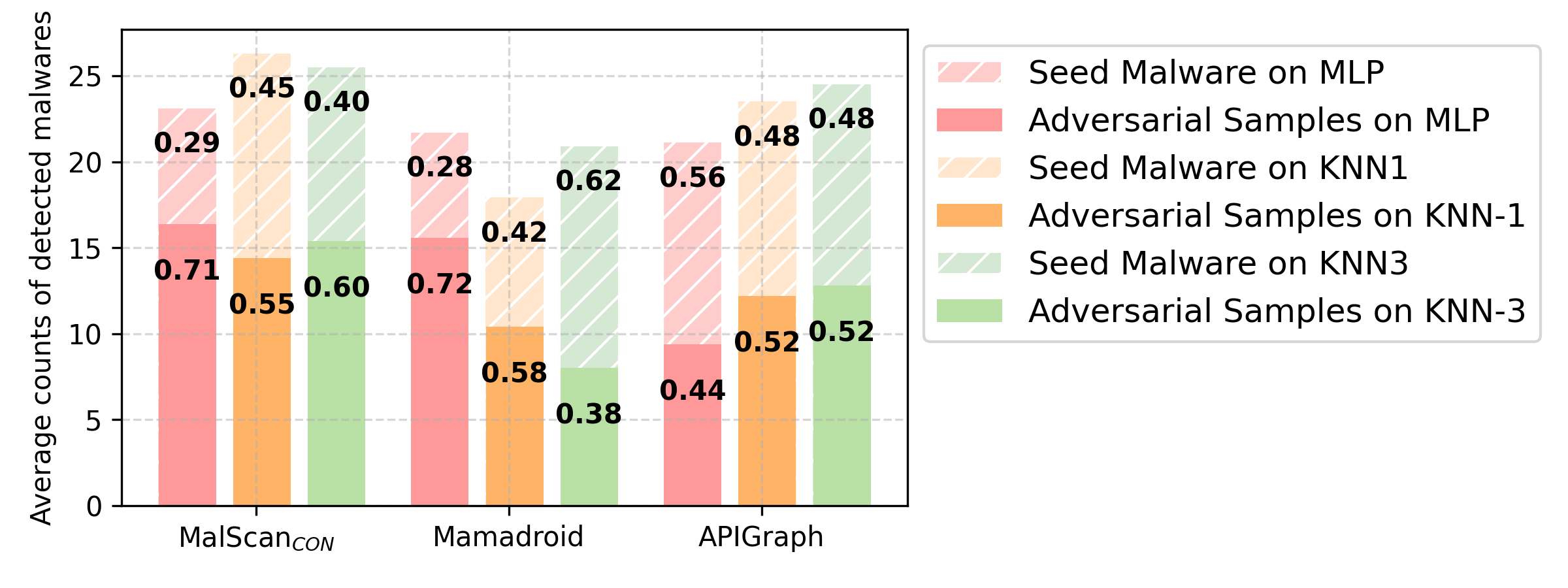}
  % \caption{Number of malware and their adversarial samples detected by VirusTotal}
  % \vspace{-15pt}
  \caption{Detection Score Difference Between Malware and Corresponding Adversarial Samples.}
  \label{fig:rq4_virustotal}
  \vspace{-10pt} % 调整到你满意的间距，这里使用-10pt作为示例
\end{figure}

\textbf{(1) Evaluation on Real-world AMD.}
% \subsubsection{Robustness in the Real-world}
% \hfill \\
% \textbf{Setup.}
% To understand the capability of \tool in real-world robustness evaluation,  we selected VirusTotal~\cite{virustotal}, a leading platform for malware analysis, as our target~\cite{li2023black,sun2023mate,he2023efficient}. 
{To explore the robustness issues present in real-world models, we selected VirusTotal~\cite{virustotal}, a leading platform for malware analysis, as our target~\cite{li2023black,sun2023mate,he2023efficient}.}
% Based on the transferability of different features and models, as detailed on our website~\cite{ourwebsite}, we try to evaluate the real-world robustness of malware detectors.
% To achieve this, we selected VirusTotal~\cite{virustotal}, a leading platform for analyzing malware, as our target~\cite{li2023black,sun2023mate,he2023efficient}. 
% Detailed experimental setup in our website.
% We randomly selected 10 adversarial samples from each scenario with original detection counts varying from 4 to 45. We analyzed how frequently these samples were detected as malware, and compared their detection scores as malware before and after the attacks.
We randomly selected 10 adversarial samples from each scenario, with original detection scores ranging from 4 to 45, and compared their detection scores before and after our attack.
{Figure~\ref{fig:rq4_virustotal} shows the change in detection scores for the original malware and their adversarial counterparts as analyzed by VirusTotal.}
%
%comparing these changes for the original samples both before and after the attacks.
%
% \noindent\textbf{Results \& Analysis.}
% Figure~\ref{fig:rq4_virustotal} shows the difference in detection scores as analyzed by VirusTotal.
% Each pair of bars represents one feature-classifier combination, with the taller bar (striped fill) for the original malware, and the shorter bar (solid fill) for the adversarial samples. The numbers inside the bars with striped fill represent the rate of change in detected samples, highlighting the proportion of samples bypassing detection engines. 
% Notably, the detection scores (i.e., degree of maliciousness) decreased significantly (over 28\%), indicating potential robustness issues in the real-world AMD of several vendors, possibly due to reliance on vulnerable detection algorithms.
{Notably, the detection scores (which indicate the level of maliciousness) dropped significantly (by over 28\%), suggesting potential weak robustness in the real-world AMD systems of several vendors. This drop might be due to these systems relying on detection algorithms that have robustness issues.}
% adversarial samples successfully bypassed some antivirus mechanisms.
% This suggests the robustness issues in the malware detection methods employed by at least one-third of the vendors, possibly due to reliance on outdated or simplistic detection algorithms. 
% This indicates potential robustness issues in the real-wolrd AMD of several vendors, possibly due to reliance on outdated or simplistic detection algorithms.
Further analysis shows that the impact of adversarial samples varies across different classifiers. 
KNN-3 experiences the greatest impact (average 50\%) from a model perspective, while \apigraph has the greatest impact (average 51\%) from a feature perspective.
This suggests a preference for relatively robust features and models in real-world scenarios, but they still cannot withstand strong adversarial attacks (e.g., \tool). 
To further understand why \tool performs effectively in real-world black-box systems (i.e., VirusTotal), we analyzed the transferability across different features and models. For more detailed results, please refer to our website~\cite{ourwebsite}.
% Therefore, we can identify these vulnerabilities to help improve the robustness.

% On average, the effects are 0.38 for MLP, 0.45 for KNN-1, and 0.50 for KNN-3, indicating that AV engines possibly using KNN-like mechanisms based on sample similarity and voting are particularly vulnerable. Furthermore, the impact of adversarial samples increased with the abstraction level of features, where the average impact is 0.38 for \malscan, 0.44 for \mamadroid, and 0.51 for \apigraph, reflecting its preference in real-world setups due to its detailed API-based semantic analysis. According to Finding \#3, \apigraph is considered a relatively robust feature, but it is still vulnerable under \tool's attack. Therefore, this suggests that combining robust features with strong models (e.g., RF), is crucial for enhancing resilience against diverse attacks in real-world scenarios.

% \vspace{-0.15cm}
% \begin{tcolorbox}[left=1pt, right=1pt, top=0pt, bottom=0pt]
% \label{conclusion:4}
% \small
% \hypertarget{}
% {\textbf{Answer to RQ4}}: 
% \textit{
% The results demonstrate that the generated adversarial samples exhibit transferability across different models and features. This characteristic is particularly useful for evaluating real-world black-box antivirus engines.
% }
% \end{tcolorbox}

\noindent\textbf{(2) Robustness Enhancement via Retraining.}
% Robustness concerns in machine learning have long been a significant area of study~\cite{sehwag2019analyzing, lee2021machine}. 
The experimental results reinforce the need for continued enhancements in AMD robustness. 
From the feature perspective, integrating robust features such as MalScan (Katz) and \apigraph to comprehensively represent malware behaviors proves to be a promising approach.  This method, akin to merging static and dynamic features, leverages the distinct advantages of each to provide a complete depiction of potential threats~\cite{anupama2022detection}.
From the model perspective, retraining has been a popular strategy to boost ML model robustness~\cite{chen2020explore, mani2021defending, chen2023continuous}. 
In supplementary experiments, we retrained models with adversarial examples generated by \tool, which could significantly decrease the ASR (see detailed results on the website~\cite{ourwebsite}).
% Our experimental results further affirm this approach's efficacy, particularly when incorporating adversarial examples generated by \tool (detailed in Appendix~\ref{retrain}). 
Additionally, our studies suggest that ensemble models serve as more robust classifiers in FCG-based AMD environments.

\noindent\textbf{(3) High-efficiency Strategies for Ensemble Models.}
The perturbation rates for ensemble models were observably high (\S \ref{evaluation:RQ1 effectiveness}), partially due to a limited number of modifiable nodes (i.e., user function calls) in certain malware. This led us to propose the new node-based mutation operators (i.e., ASN and ADN). However, we found that their effectiveness was still limited to specific scenarios. These results indicate that the mutation operators could still be improved, especially by developing edge-based perturbation variants to potentially increase effectiveness on ensemble models, as demonstrated by the success of edge-based ALE in \S \ref{evaluation:RQ3 ablation}.
Moreover, we can also attempt to develop some interpretation-based feedback at the operator level, which is capable of identifying more influential and less frequently altered operations.

% \end{enumerate}

% Robustness issues in ML have long been a focal point of concern~\cite{sehwag2019analyzing, lee2021machine}. Our research shows that enhancing AMD robustness is still necessary. From the feature perspective, we can combine the more robust MalScan\textsubscript{K} and \apigraph to represent malware behaviors. This approach is similar to integrating static and dynamic features, leveraging the unique strengths of each to provide a comprehensive view of potential threats~\cite{anupama2022detection}.
% From the model perspective, retraining is a common method to enhance the robustness of ML models~\cite{chen2020explore, mani2021defending}, and our experiments also demonstrate its usefulness by adding the adversarial examples generated by \tool (see details in the Appendix~\ref{retrain}). Moreover, our findings indicate that ensemble models are more robust classifiers in FCG-based AMD. 
% \vspace{-0.25cm} 
\section{Threats to Validity}\label{threatstovalidity}
% \vspace{-0.15cm} 
% \textbf{(1) Limited Malware Seeds.}\feng{Better not mention HRAT specifically. Overlapped with (3)}
%由于hrat的运行速度非常的慢，因此我们只选择了有限的数据集在实验中，但是为了确保数据有效且具有代表性，我们分别从6年（覆盖最新年份）中均匀的选择seed，并且确保了这些seed的size是不同的，确保多样化。
The selection of the dataset, including the training dataset and the seed malware, poses a threat to validity. We address this threat by adhering to established data selection protocols and collecting a diverse range of APKs from 2018 to 2023. Similarly, to ensure the validity and representativeness of the seed malware, we randomly selected these seeds from the past six years and varied their sizes to maintain diversity.
{To the best of our knowledge, our dataset spans a recent six-year range, providing broader coverage compared to the baselines.}

The selection of models presents another potential threat to validity, as the results may vary in different AMD models. To mitigate it, we have endeavored to include a broad range of categories, incorporating features with various granularity from the FCG and multiple machine learning models with distinct decision mechanisms. To the best of our knowledge, our evaluations are the most comprehensive concerning various FCG-based features and models.

The replication and extension of baselines pose another threat to validity. Notably, significant discrepancies persist between our results and those reported in the original papers. Actually, the reproducibility issues have also been acknowledged by existing works~\cite{olszewski2023get}. We addressed this threat carefully by: 1) meticulously reviewing the code and consulting with their authors to clarify ambiguous parts; 2) engaging in discussions with the authors about the discrepancies, attributing potential causes to differences in datasets, the APK extraction tools used (e.g., Androguard~\cite{androguard} versus FlowDroid~\cite{flowdroid}), and the models evaluated; 3) releasing our code, models, and seed malware to facilitate verification and replication of our findings~\cite{ourwebsite}.

Finally, employing the interpretation-based method (i.e., SHAP) for interpreting model decisions could pose a threat to validity. SHAP values may not always accurately reflect the influence of different inputs in models. Additionally, alternative interpretation methods could be considered. To mitigate this, we did not rely solely on interpretation scores; instead, the model's output served as the primary and dominant feedback. Our extensive evaluation also demonstrates the overall usefulness of this approach. In future work, we plan to explore the impact of various interpretation methods on \tool.

%% file: tex/7-related_work.tex
\section{Related Work}
% \vspace{-0.15cm}
% \textbf{Android Malware Detection.}\cs{here} 

% ML techniques have gained significant traction in the domain of AMD,  leveraging diverse feature extraction methodologies and ML classifiers to determine whether an APK is benign or malicious. 
% Early works~\cite{arp2014drebin,avdiienko2015mining,garcia2018lightweight, bai2020unsuccessful, bai2021comparative} focus on string-based detectors, like Drebin~\cite{arp2014drebin}, use one-hot encoding to represent the usage of critical APIs. However, string-based approaches often lack deeper semantic information needed to accurately represent malicious patterns, prompting a shift towards more sophisticated techniques such as code structure representations through graph~\cite{wu2019malscan,onwuzurike2019mamadroid,zhang2020enhancing, gao2024comprehensive}.
% Graph-based detectors, on the other hand, use various metrics to represent the calls of critical APIs, offering improved robustness and efficacy. 
% For example, \malscan~\cite{wu2019malscan} represents APK behaviors by the importance of critical nodes (system APIs) from a function-level call graph.

% Related works has primarily focused on various techniques for generating adversarial examples for evaluating their robustness. 
%
%
%在String-based上的adversarial attack是非常容易的，由于他们是one-hot编码，利用gradient-based的方法对feature进行修改之后，可以跟code一一对应。而在graph-based上的对抗攻击，存在problem-feature reverse的问题，一些work专注于feature level但是难以保持apk的功能完整性，近期的工作专注于在code level进行探索，并且利用启发式搜索的算法，例如强化学习、遗传算法等。
%
{\textbf{ML-based Android Malware Detection.}
ML techniques have gained significant traction in the domain of Android malware detection, leveraging diverse feature extraction and embedding methodologies, such as string-based~\cite{arp2014drebin,garcia2018lightweight}, image-based~\cite{xiao2019image,yuan2020byte}, graph-based~\cite{wu2019malscan,onwuzurike2019mamadroid,zhang2020enhancing}. 
For instance, Drebin~\cite{arp2014drebin} utilizes static strings such as permissions and API calls extracted from APKs, employing Support Vector Machines (SVM) for classification. Addressing string obfuscation, RevealDroid~\cite{garcia2018lightweight} resorts to byte-code extraction for consistent classification with Drebin.
However, string/image-based approaches often lack semantic information, prompting a shift towards more sophisticated techniques like Function Call Graph (FCG) representations, exemplified by \malscan~\cite{wu2019malscan}. \malscan represents FCG from the smali code as a social network and employs k-nearest neighbors (KNN) as the classifier, offering improved robustness and efficacy.}

{\textbf{Adversarial Attacks for Robustness Evaluation.}
% Related works~\cite{pierazzi2020intriguing,chen2019android,li2020adversarial,zhao2021structural,he2023efficient,li2023black} in the field of adversarial attacks on ML/DL-based AMD methods has primarily focused on various techniques for generating adversarial examples and evaluating their effectiveness. 
% For instance, APG~\cite{pierazzi2020intriguing} employs gradient-based searches, while Android HIV~\cite{chen2019android} targets logits layers using algorithms like C\&W. Recent advances have shifted towards attacks from the problem space by maintaining the semantic integrity of samples, thereby preserving their operational viability in practical scenarios~\cite{zhao2021structural,he2023efficient,li2023black}.
% For example, \hrat~\cite{zhao2021structural} propose graph modification rules for reinforcement learning-based attacks against KNN-1, while \bagammo~\cite{li2023black} focus on GA-based attacks when embedding methods are unknown, primarily targeting \mamadroid~\cite{onwuzurike2019mamadroid}. However, these approaches often rely heavily on model gradient information and may not generalize well across different embedding methods or target models.
Related works~\cite{pierazzi2020intriguing,chen2019android,li2020adversarial,zhao2021structural,he2023efficient,li2023black} have primarily focused on various techniques for generating adversarial examples to evaluate the robustness of malware detectors. 
Abundant adversarial attacks on string-based detectors are relatively simple and straightforward features using one-hot encoding, like Drebin~\cite{arp2014drebin}.
By using gradient-based methods~\cite{pierazzi2020intriguing, chen2019android, zhang2021shadowdroid} or interpretability-assisted techniques~\cite{severi2021explanation, sun2023mate}, features can be directly modified and mapped back to the code, leading to high attack success rates.
However, adversarial attacks on graph-based detectors face the problem-feature reverse challenge. This work~\cite{chen2019android} focused on the feature level but struggled to maintain the functional integrity of the APK. 
Two recent studies have shifted towards exploring code-level attacks, employing heuristic search algorithms~\cite{zhao2021structural, li2023black}.
{
Zhao et al.~\cite{zhao2021structural} exploited gradient information to estimate perturbation locations and directions. However, discrete gradient estimation errors on binary graphs may cause reinforcement learning to proceed in the wrong direction.
Li et al.~\cite{li2023black} utilized single perturbations and scores from a surrogate model to guide the attack process. However, it easily falls into local optima due to the vast perturbation space created by sparse features like \malscan, resulting from a lack of precise feedback and diversified operators.} 
Other researchers are exploring more efficient adversarial attack techniques to improve robustness evaluations of AMD methods, such as using interpretation-assisted feedback~\cite{amich2022eg, liu2022explanation, sun2023mate,yu2023airs}. For example,
%他利用可解释性的引导产生对ML模型有意义的对抗例子，寻找有意义的扰动，帮助评估系统的鲁棒性。
Amich et al.~\cite{amich2022eg} leverages SHAP to guide 
adversarial example crafting against ML models, seeking meaningful perturbations to aid in assessing the system's robustness.
Sun et al.\cite{sun2023mate} utilize SHAP to guide attacks on string-based detectors, identifying critical API permissions and inserting uncalled functions. Similarly, Yu et al.\cite{yu2023airs} propose step-level interpretability feedback for deep reinforcement learning in security, aiding in identifying critical steps.}
%Inspired by these research efforts, we aim to leverage interpretation-assisted adversarial attacks to evaluate the robustness of FCG-based AMD methods.

{
Compared to these studies, which focus on proposing adversarial attacks, we concentrate on testing the robustness of graph-based malware detectors through adversarial graph/sample generation and providing findings to enhance robustness.
% Specially, our approach uses novel strategies, such as multi-objective feedback, to guide models more effectively for evaluating the robustness of FCG-based AMD methods. 
Additionally, our method does not require the model to be differentiable and significantly broadens the range of target features and models.
}

% Recent advances have shifted towards attacks from the problem space by maintaining the semantic integrity of samples, thereby preserving their operational viability in practical scenarios~\cite{zhao2021structural,he2023efficient,li2023black}.
% For example, \hrat~\cite{zhao2021structural} propose graph modification rules for reinforcement learning-based attacks against KNN-1, while \bagammo~\cite{li2023black} focus on GA-based attacks when embedding methods are unknown, primarily targeting \mamadroid~\cite{onwuzurike2019mamadroid}. However, these approaches often rely heavily on model gradient information and may not generalize well across different embedding methods or target models.

% Other researchers are exploring more efficient adversarial attack techniques to improve robustness evaluations of AMD methods, such as using interpretation-assisted feedback~\cite{amich2022eg, liu2022explanation, sun2023mate,yu2023airs}. 
% For example, Amich et al.~\cite{amich2022eg} leverages SHAP to guide 
% adversarial example crafting against ML models, seeking meaningful perturbations to aid in assessing the system's robustness.
% Sun et al.\cite{sun2023mate} utilize SHAP to guide attacks on string-based detectors, identifying critical API permissions and inserting uncalled functions.
% Similarly, Yu et al.\cite{yu2023airs} propose step-level interpretability feedback for deep reinforcement learning in security, aiding in identifying critical steps.

%% file: tex/9-conclusion.tex
% \vspace{-0.2cm} 
\section{Conclusion}
% \vspace{-0.15cm} 

In this paper, we introduce a method to evaluate the robustness of FCG-based AMD systems. 
This method incorporates dependency-aware mutation strategies and utilizes innovative interpretation-based fitness functions to effectively guide perturbation optimization within an FCG. 
Our experiments demonstrate superior performance across diverse 40 scenarios, and achieve an average attack success rate of 87.9\%, significantly outperforming baseline methods. Furthermore, our findings offer valuable insights for enhancing model robustness in future developments, and we also provide an in-depth discussion on the benefits of adversarial retraining.

% % \vspace{-0.25cm} 
% \section{Data Availability}
% % \vspace{-0.15cm} 
% The datasets and code used in this study are available in our anonymized repository~\cite{ourcode}.